\documentclass[12pt]{article}

\usepackage[english]{babel}
\usepackage[usenames]{color}
\usepackage[cp1250]{inputenc}
\usepackage{amsfonts}
\usepackage{amsthm}
\usepackage{graphicx}
\usepackage{epsfig}

\theoremstyle{plain}

\begin{document}
\newcommand{\bea}{\begin{eqnarray}}
\newcommand{\eea}{\end{eqnarray}}
\newcommand{\be}{\begin{equation}}
\newcommand{\ee}{\end{equation}}
\newcommand{\beas}{\begin{eqnarray*}}
\newcommand{\eeas}{\end{eqnarray*}}
\newcommand{\bs}{\backslash}
\newcommand{\bc}{\begin{center}}
\newcommand{\ec}{\end{center}}

\title{Asymmetric numeral systems.}

\author{Jarek Duda}

\date{\it \footnotesize Jagiellonian University,  Poland,
\textit{email:} dudaj@interia.pl}

\maketitle

\begin{abstract}
In this paper will be presented new approach to entropy coding:
family of generalizations of standard numeral systems which are
optimal for encoding sequence of equiprobable symbols, into
asymmetric numeral systems - optimal for freely chosen probability
distributions of symbols. It has some similarities to Range Coding
but instead of encoding symbol in choosing a range, we spread
these ranges uniformly over the whole interval. This leads to simpler
encoder - instead of using two states to define range, we need
only one. This approach is very universal - we can obtain from
extremely precise encoding (ABS) to extremely fast with
possibility to additionally encrypt the data (ANS). This
encryption uses the key to initialize random number generator,
which is used to calculate the coding tables. Such preinitialized
encryption has additional advantage: is resistant to brute force
attack - to check a key we have to make whole initialization. There will
be also presented application for new approach to error correction: after an error
in each step we have chosen probability to observe that something was wrong.
We can get near Shannon's limit for any noise level this way
with expected linear time of correction.
\end{abstract}

\section{Introduction}
In practice there are used two approaches for entropy coding
nowadays: building binary tree (Huffman coding \cite{huf}) and
arithmetic/range coding (\cite{ari},\cite{ran}). The first one
approximates probabilities of symbols with powers of 2 - isn't
precise. Arithmetic coding is precise. It encodes symbol in
choosing one of large ranges of length proportional to assumed
probability distribution ($q$).  Intuitively, by analogue to
standard numeral systems - the symbol is encoded on the most
important position. To define the current range, we need to use two numbers (states).\\

We will construct precise encoding that uses only one state. It will be done by
distributing symbols uniformly instead of in ranges -
intuitively: place information on the least important position.
Standard numeral systems are optimal for encoding streams of
equiprobable digits. Asymmetric numeral systems (\cite{me}) is natural
generalization into other, freely chosen probability distributions.
If we choose uniform probability, with proper initialization we
get standard numeral system.\\

For the binary case: Asymmetric Binary System (ABS) there are found
practical formulas, which gives extremely precise entropy encoder
for which probability distribution of symbols can freely change. It
was show (\cite{mah}) that it can be practical alternative for
arithmetic coding.

For the general case: Asymmetric Numeral Systems (ANS) instead of
using formulas, we initially use pseudorandom number generator to
distribute symbols with assumed statistics. The precision can be
still very high, but disadvantage is that when the probability
distribution changes, we have to reinitialize. The advantage is
that we encode/decode a few bits in one use of the table - we get
compression rates like in arithmetic coding and transfers like in
Huffman coding. On \cite{dem} is available demonstration.

Another advantage is that we can use a key as the
initialization of the random number generator, additionally
encrypting the data. Such encryption is extremely unpredictable - uses random coding tables and hidden random variable to choose behavior and so the current length of
block. This approach is faster than standard block ciphers and is much more resistant against
brute force attacks. \\

In the last section will be presented new approach to error correction, which
is able to get near Shannon's limit for any noise level and is still practical - has
expected linear (or $N\lg(N)$) correction time. It can be imagined
as path tracking - we know starting and ending position and we want to walk between
them using the proper path. When we use this path everything is fine, but when we lost
it, in each step we have selected probability of becoming conscious of this fact. Now we can go back and try to make some correction. If this probability is chosen higher than
some threshold corresponding to Shannon's limit, the number of corrections we should try
doesn't longer grow exponentially and so we can easily verify that it was the proper correction.

Intuitively we use short blocks, but we connect their redundancy. This connection practically allows to 'transfer' surpluses of redundancy to help with large local error concentrations.
\subsubsection{Very brief introduction to entropy coding}
In the possibility of choosing one of $2^n$ choices is stored $n$
bits of information. Assume now that we can store information in
choosing a sequence of bits of length $n$, but such that the
probability of '1' is given ($p$). We can evaluate the number of
such sequences using Stirling's formula
$\left(\lim_{n\to\infty}\frac{n!}{\sqrt{2\pi
n}\left(\frac{n}{e}\right)^n}=1\right)$:
\begin{eqnarray*}
{n \choose pn}&=& \frac{n!}{(pn)!(\tilde{p}n)!}\approx
(2\pi)^{-1/2}\frac{n^{n+1/2}e^n}{(pn)^{pn+1/2}(\tilde{p}n)^{\tilde{p}n+1/2}e^n}=\\&=&
(2\pi np\tilde{p})^{-1/2}p^{-pn}\tilde{p}^{-\tilde{p}n}=(2\pi
np\tilde{p})^{-1/2}2^{-n(p\lg p+\tilde{p} \lg{\tilde{p}})}
\end{eqnarray*}
where $\tilde{p}=1-p$. So while encoding in such sequences, we can store at average \be
\label{entr} h(p):=-p\lg(p)-(1-p)\lg(1-p)\quad \textrm{bits of
information/symbol}\ee That's well known formula for Shannon's
entropy. In practice we usually don't know the probability
distribution, but we are approximating it using some statistical
analysis. The nearer it is to the real probability distribution,
the better compression rates we get. The final step is the entropy
coder, which uses found statistics to encode the message.\\

Even if we would know the probability distribution perfectly, the
expected compression rate would be usually a bit larger than
Shannon's entropy. One of the reason is that encoded message usually contains
some additional correlations. The second
source of such entropy increase is that entropy coders are
constructed for some discrete set of probability distributions, so
they have to approximate the original one.

In an event of probability $1/n$ is stored $\lg(n)$ bits, so
generally in event of probability $q$, should be stored $\lg(1/q)$
bits. This can be seen in Shannon's formula: it's average of
stored bits with probabilities of events as weights.

So if we use a coder which encodes perfectly $(q_s)$ symbol distribution to
encode $(p_s)$ symbol sequence, we would get at average $\sum_s
p_s \lg(1/q_s)$ bits per symbol. The difference between this value
and the optimal one is called Kullback - Leiber distance: \be
\Delta H=\sum_s p_s \lg\left(\frac{p_s}{q_s}\right)\approx
\sum_s\frac{-p_s}{\ln(2)}\left(\left(1-\frac{q_s}{p_s}\right)-
\frac{1}{2}\left(1-\frac{q_s}{p_s}\right)^2\right)\approx
0.72\sum_s \frac{(p_s-q_s)^2}{p_s}\label{kld}\ee

We have used second order Taylor's expansion of logarithm around
1. The first term vanishes and the second allows to quickly
estimate how important is that entropy coder is precise.

\section{General concept} We would like to encode an
uncorrelated sequence of symbols of known probability distribution
into as short as possible sequence of bits. For simplicity we will
assume that the probability distribution doesn't change in
time, but it can be naturally generalized to varying
distributions. The encoder will receive succeeding symbols and
transform them into succeeding bits.\\

 An symbol(event) of probability $p$ contains
$\lg(1/p)$ bits of information - it doesn't have to be a natural
number. If we just assign to each symbol a sequence of bits like
in Huffman coding, we approximate probabilities with powers of 2.
If we want to get closer to the optimal compression rates, we have
to be more precise - the encoder have to be
more complicated - use not only the current symbol, but also
relate for example to the previous ones. The encoder should have
some state in which is stored unnatural number of bits of
information. This state in arithmetic coder are two numbers
describing the current range.\\

The state of presented encoder will be one natural number:
$x\in\mathbb{N}$.  For this subsection we will forget about
sending bits to output and focus on encoding symbols. So the state
$x$ in given moment is a large natural number which encodes all already
processed symbols. We could just encode it as a binary number
after processing the whole sequence, but because of its size it's
completely impractical. In section 4 it will be shown that we can
transfer the youngest bits of $x$ to assure that it stays in some
fixed range during the whole process. For now we are looking for a
rule of changing the state while processing a symbol $s$:
\begin{equation}
 (s,x)\begin{array}{cccc}\textrm{encoding}\\\longrightarrow\\\longleftarrow\\\textrm{decoding}
 \end{array}x'
\end{equation}
So our encoder starts with for example $x=0$ and uses above rule on
succeeding symbols. These rules are bijective, so that we can
uniquely reverse whole process - decode the final state back into
initial sequence of symbols in reversed order.

In given moment in $x$ is stored some unnatural number of bits of
information. While writing it in binary system, we would round
this value up. To avoid such approximations, we will use
convention that $x$ is the
possibility of choosing one of $\{0,1,..,x-1\}$ numbers, so $x$ contains exactly $\lg(x)$ bits of information.\\

For assumed probability distribution of $n$ symbols, we will
somehow split the set $\{0,1,..,x-1\}$ into $n$ separate subsets -
of sizes $x_0,..,x_{n-1}\in\mathbb{N}$, such that
$\sum_{s=0}^{n-1} x_s=x$. We can treat the possibility of choosing
one of $x$ numbers as the possibility of choosing the number of
subset($s$) and then choosing one of $x_s$ numbers. So with
probability $q_s=\frac{x_s}{x}$ we would choose $s$-th subset. We
can enumerate elements of $s$-th subset from $0$ to $x_s-1$ in the
same order as in the original enumeration of $\{0,1,..,x-1\}$.

Summarizing: we've exchanged the possibility of choosing one of
$x$ numbers ($\lg(x)$ bits) into the possibility of choosing a
pair: a symbol $s$ ($lg(1/q_s)$ bits) with known probability
($q_s$) and the possibility of choosing one of $x_s$
numbers ($\lg(x_s)=\lg(x)-\lg(q_s)$ bits). This ($x
\rightleftharpoons (s,x_s)$) will be the bijective coding we are
looking for.

We will now describe how to split the range. In arithmetic coding
approach (Range Coding), we would divide $\{0,..,x-1\}$ into
ranges. In ANS we will distribute these subsets
uniformly.\\

We can describe this split using \textbf{distributing function}
$D_1:\mathbb{N}\to\{0,..,n-1\}$:
$$\{0,..,x-1\}=\bigcup_{s=0}^{n-1} \{y\in\{0,..,x-1\}:D_1(y)=s\}$$
We can now enumerate numbers in these subsets by counting how many
elements from the same subset was there before: \be \label{C2}
x_s:=\#\{y\in\{0,1,..,x-1\},\ D_1(y)=s\}\quad\quad\quad
D_2(x):=x_{D_1(x)}\ee getting bijective \textbf{decoding
function}(D) and it's inverse \textbf{coding function} (C):
$$D(x):=(D_1(x),D_2(x))=(s,x_s)\qquad \qquad C(s,x_s):=x.$$

Assume that our sequence consists of $n\in\mathbb{N}$ symbols with
given  probability distribution $(q_s)_{s=0,..,n-1}$ $\
\left(\forall_{s=0,..,n-1}\ q_s>0\right)$. We have to construct a
distributing function and coding/decoding function for this
distribution: such that \be\forall_{s,x}\quad\quad x_s \ \
\textrm{is approximately}\ \ x\cdot q_s\ee We will now show
informally how essential above condition is. In section 3 and 5 will
be shown two ways of making such construction.\\

Statistically in a symbol is encoded $H(q):=-\sum_s q_s \lg{q_s}$
bits.\\ ANS uses $\lg(x)-\lg(x_s)=\lg(x/x_s)$ bits of information
to encode a symbol $s$ from $x_s$ state. Analogously to (\ref{kld}) using second Taylor's
expansion of logarithm (around $q_s$), we can estimate that our
encoder needs at average:
$$-\sum_s q_s \lg\left(\frac{x_s}{x}\right)\approx -\sum_s q_s
\left(\lg(q_s)+\frac{x_s/x-q_s}{q_s\ln(2)}-\frac{(x_s/x-q_s)^2}{2q_s^2\ln(2)}\right)=$$
$$=H(q)+\frac{1-1}{q_s\ln(2)}+\sum_s \frac{(x_s/x-q_s)^2}{2q_s\ln(2)}\ \ \ \ \textrm{bits/symbol.}$$
We could average \be\label{acc} \frac{1}{2\ln(2)}\sum_s
\frac{q_s}{x^2}(x_s/q_s-x)^2 =\frac{1}{\ln(4)}\sum_s
\frac{q_s}{x^2}\left(x_s/q_s-C(s,x_s)\right)^2 \ee over all
possible $x_s$ to estimate how many bits/symbols we are wasting. We will do it in section 6.
\section{Asymmetric Binary System (ABS)}
 It occurs that in the binary case we can find simple explicit
formula for coding/decoding functions.\\

We have now two symbols: $"0"$ and $"1"$. Denote $q:=q_1$, so $\tilde q:=1-q=q_0$.\\
To get $x_s\approx x\cdot q_s$, we can for example take \be
x_1:=\lceil xq
\rceil\quad\quad\quad\quad\quad\quad\left(\textrm{or
alternatively}\ x_1:=\lfloor xq \rfloor\right)  \ee \be
x_0=x-x_1=x-\lceil xq \rceil
\quad\quad\quad\quad\quad\left(\textrm{or}\ x_0=x-\lfloor xq
\rfloor \right)\ee Now using (\ref{C2}): $D_1(x)=1\
\Leftrightarrow$ there is a jump of $\lceil xq \rceil$ after it:
\be s:=\lceil (x+1)q \rceil-\lceil xq \rceil
\quad\quad\quad\left(\textrm{or}\ s:=\lfloor (x+1)q
\rfloor-\lfloor xq \rfloor\right) \ee
We've just defined \textbf{decoding} function: $D(x)=(s,x_s)$.\\

For example for $q=0.3$:\\

\begin{tabular}{|c||c|c|c|c|c|c|c|c|c|c|c|c|c|c|c|c|c|c|c|}\hline
$x$&0&1&2&3&4&5&6&7&8&9&10&11&12&13&14&15&16&17&18
\\\hline\hline $x_0$&&0&1&&2&3&&4&5&6&&7&8&&9&10&&11&12\\\hline
$x_1$&0&&&1&&&2&&&&3&&&4&&&5&&\\\hline
\end{tabular}\\
 \\

We will find coding function now: we have $s$ and $x_s$ and want to find $x$.\\
Denote $r:=\lceil xq \rceil-xq\in[0,1)$\\
$s:=\lceil (x+1)q \rceil-\lceil xq \rceil=\lceil (x+1)q -\lceil xq
\rceil\rceil=\lceil (x+1)q-r-xq\rceil=\lceil q-r \rceil$\\
$$s=1\Leftrightarrow r<q$$
\begin{itemize}
\item $s=1$: $\quad x_1=\lceil xq \rceil=xq+r$\\
$x=\frac{x_1-r}{q}=\Big\lfloor\frac{x_1}{q}\Big\rfloor\quad$
because it's natural number and $0\leq r<q$.
\item $s=0$: $q\leq r<1$ so $\tilde{q}\geq 1-r>0$\\
$x_0=x-\lceil xq \rceil=x-xq-r=x\tilde{q}-r$
$$x=\frac{x_0+r}{\tilde{q}}=\frac{x_0+1}{\tilde{q}}-\frac{1-r}{\tilde{q}}=
\Big\lceil\frac{x_0+1}{\tilde{q}}\Big\rceil-1$$
\end{itemize}
Finally \textbf{coding}: \be \label{decoding} C(s,x)=\left\{
\begin{array}{ll}
\Big\lceil\frac{x+1}{1-q}\Big\rceil-1 &\ \textrm{if}\ s=0\\
\ \Big\lfloor\frac{x}{q}\Big\rfloor &\ \textrm{if}\ s=1\end{array}
\right. \quad\quad\quad\left(\textrm{or}\ = \left\{
\begin{array}{ll}
\ \ \ \Big\lfloor\frac{x}{1-q}\Big\rfloor &\ \textrm{if}\ s=0\\
\Big\lceil\frac{x+1}{q}\Big\rceil-1 &\ \textrm{if}\ s=1\end{array}
\right.\right)\ee For $q=1/2$ it's usual binary system (with
switched digits).
\section{Stream coding/decoding} We can
encode now into a large natural numbers ($x$). We would like to use
ABS/ANS to encode data stream - into potentially infinite sequence
of digits(bits) with expected uniform distribution. To do it we
can sometimes transfer a part of information from $x$ into a digit
from a standard numeral system to enforce $x$ to
stay in some fixed range ($I$).\\

\subsection{Algorithm}
Let us choose that the data stream will be encoded as $\{0,..,b-1\}$ \emph{digits} - in
standard numeral system of base $b\geq 2$. In practice we should mainly use
the binary system ($b=2$), but thanks of this general approach, we can
for example use $b=2^8$ to transfer whole byte at once. Symbols
contain correspondingly $\lg(1/q_s)$ bits of information. When
they cumulate into $\lg b$ bits, we will transfer full digit
to/from output, moving $x$ back to $I$ (\emph{bit transfer}).\\

Observe that taking interval in form ($l\in\mathbb{N}$): \be
I:=\{l,l+1,..,bl-1\}\ee for any $x\in\mathbb{N}$ we have exactly
one of three cases:
\begin{itemize}
  \item $x\in I$ or
  \item $x>bl-1$, then $\exists!_{k\in\mathbb{N}}\ \lfloor x/b^k\rfloor\in
  I$ or
  \item $x<l$, then $\forall_{(d_i)\in \{0,..,b-1\}^\mathbb{N}}\
  \exists!_{k\in\mathbb{N}}\ xb^k+d_1 b^{k-1}+..+d_k\in I$.
\end{itemize}
We will call such intervals \textbf{$b$-unique}: starting from
any natural number $x$, after eventual a few reductions
($x\to\lfloor x/b\rfloor$) or placing a few youngest digits in $x$
($x\to
xb+d_t$) we would finally get into $I$ in unique way.\\

For some interval($I$), define \be I_s=\{x:C(s,x)\in I\},\quad
\textrm{so that}\ I=\bigcup_s C(s,I_s).\ee Define:\\
\begin{tabular}{l|l}
  \textbf{Stream decoding}: & \textbf{Stream coding}(\verb"s"):\\
\verb"{(s,x)=D(x);"
&  \verb"{while(x"$\notin I_s$\verb")"\\
\verb" use s;  "(e.g. to generate symbol)
&  \verb"   {put mod(x,b) to output; x="$\lfloor$\verb"x/b"$\rfloor$\verb"}"\\
\verb" while(x"$\notin I$\verb")"
& \verb" x=C(s,x)"\\
\verb"     x=xb+'digit from input'" & \verb"}"\\
\verb"}"
\end{tabular}

\begin{figure}[h]
    \centering
        \includegraphics{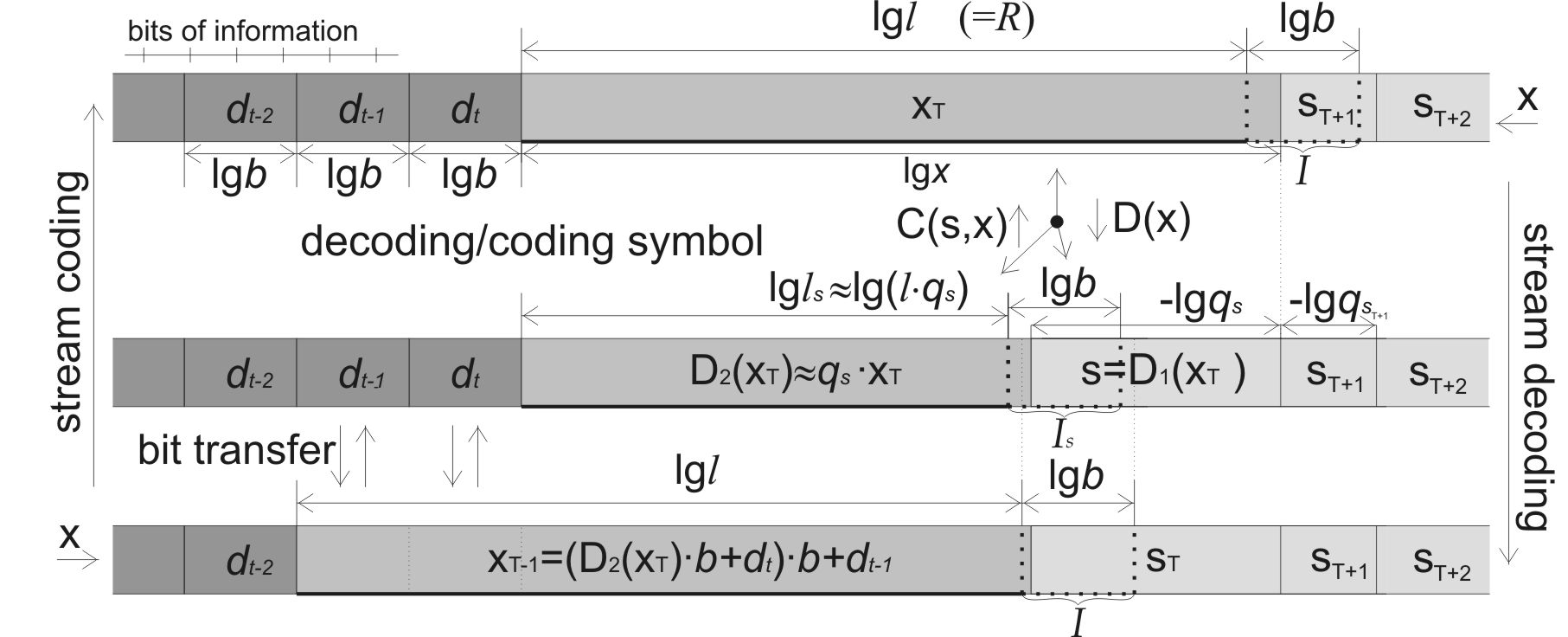}
        \caption{Stream coding/decoding}
\end{figure}

We need that above functions are ambiguous reverses of each other.\\
Observe that we would have it iff $I_s$ for $s=0,..,n-1$ and $I$
are $b$-unique: \be I=\{l,..,lb-1\}\quad\quad I_s=\{l_s,..,l_s
b-1\} \ee
 for some $l,l_s\in \mathbb{N}$.\\

We have: $\sum_s l_s(b-1)=\sum_s\#I_s=\# I=l(b-1)$.\\
Remembering that $C(s,x)\approx x/q_s$, we finally have: \be
l_s\approx lq_s\quad\quad \sum_s l_s=l.\ee

We will look at the behavior of $\lg x$ while stream coding $s$
now: \be \lg x\rightarrow\ \approx\ \lg x +\lg(1/q_s)\qquad
(\textrm{modulo}\ \lg(b))\ee We have three possible sources of
random behavior of $x$:
\begin{itemize}
\item we choose one of symbol (behavior) in statistical(random) way,
\item usually $\frac{\lg q_s}{\lg b}$ are irrational,
\item $C(s,x)$ is near but not exactly $x/q_s$.
\end{itemize}

It suggests that $\lg x$ should cover uniformly possible
space, what agrees with statistical simulations. That means that
the probability of visiting given state $x$ should be
approximately proportional to $1/x$. We will focus on it in section 6.

\subsection{Analysis of a single step}
Let's concentrate on a single stream coding step. Choose some $s\in \{0,..,n-1\}$. Among $l(b-1)$ states of $I=\{l,..,lb-1\}$ we have $l_s(b-1)$ appearances of symbol $s$.

While choosing $l_s$ we are approximating probabilities. So to simplify further analysis, let us assume for the rest of the paper: \be q_s=\frac{l_s}{l} \ee

Let us introduce for $q_s$ fractional and integer part: \be
c_s:=\{\log_b(q_s)\}\in[0,1)\qquad\qquad
k_s:=-\lfloor\log_b(q_s)\rfloor\in\mathbb{N}^+ \ee
\be\log_b(1/q_s)=k_s-c_s\qquad\qquad 1\leq q_s b^{k_s}=b^{c_s}<b\ee
where
$\{z\}=z-\lfloor z \rfloor$ is the fractional part.

Now if we introduce new variable:
\be y:=\log_b\left(\frac{x}{l}\right)\ee
we will have that one coding step is approximately $y\to^\approx\{y-c_s\}$.

\begin{figure}[h]
    \centering
        \includegraphics{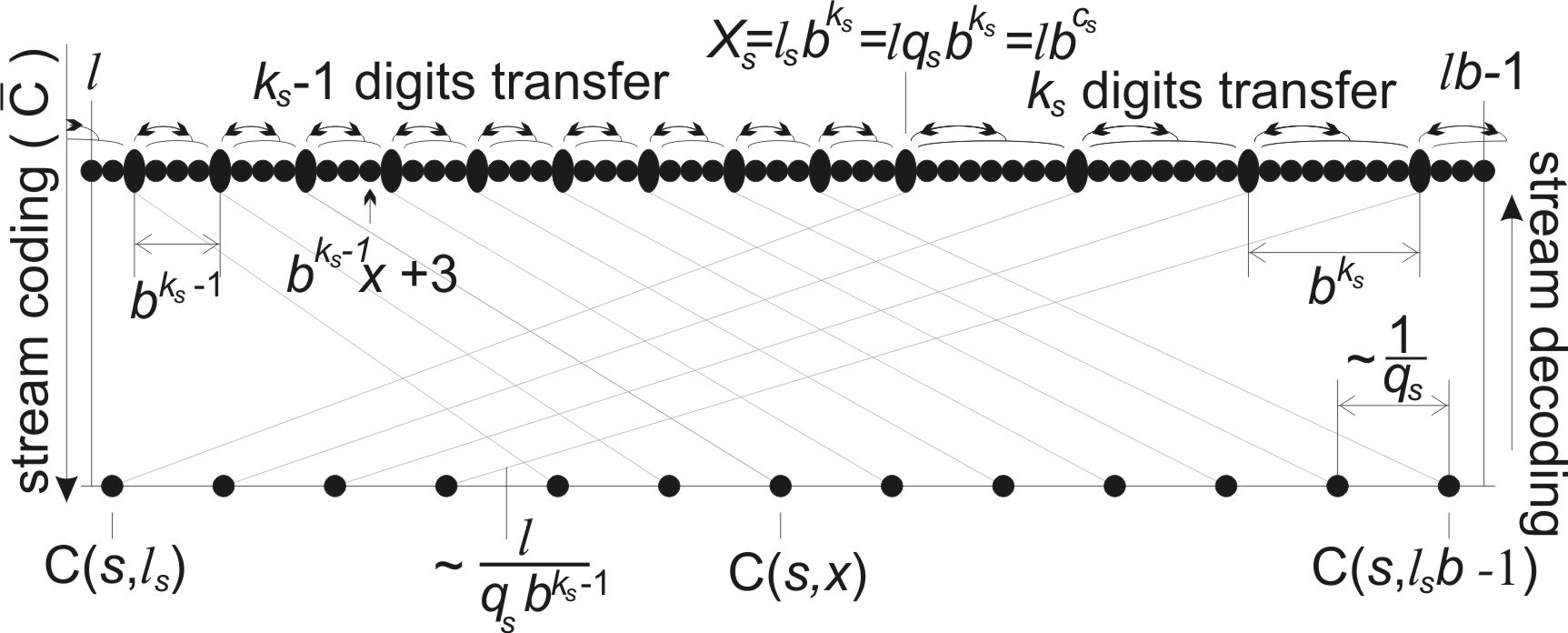}
        \caption{Example of stream
coding/decoding step for $b=2,\ k=3,\ l_s=13,\quad$  $l=9\cdot
4+3\cdot 8+6=66,\ q_s=13/66,\ x=19,\ b^{k_s-1}x+3=79=66+2+2\cdot
4+3$.}    \label{stran}
\end{figure}

The situation looks like in fig. \ref{stran}:
\begin{itemize}
  \item{The bit transfer makes that states denoted by circles will behave just
like the state denoted by ellipse on their left. The difference
between them is in transferred digits.}
\item{The number of transferred digits has maximally two
possibilities differentiating by 1: $k_s-1$ and $k_s$
$$ k_s \textrm{ is the only number such that } \lfloor
(lb-1)/b^{k_s}\rfloor \in I_s$$
  When $q_s$ is near some
integer power of $b$ ($q_s\approx b^{-k_s}$), we can have a
situation that we always transfer $k_s$ digits, but it can be
treated as a special case of the first one ($X_s=l$).}
\item{The states denoted by ellipses are multiplicities of
correspondingly $b^{k_s-1}$ or $b^{k_s}$. So if $l$ is not a
natural power of $b$, there can be some states before the first
multiplicity of $b^{k_s-1}$. They correspond to the last multiplicity
of $b^{k_s}$.

Let's assume for simplicity, that \be L:=\log_b(l)\in\mathbb{N}
\label{as1} \ee so the first state in the picture is ellipse.

With this assumption we can have special case from the previous
point, that always $k$ digits are transferred, if and only if
$q_s=b^{-k_s}$.

We assume also that we have some appearances of each symbol, so
$L\geq k_s$. }
\item{The states before the step (the top of the picture) can be
divided into two ranges - on the left or right of some boundary
value $$ X_s:=\max\{x:C(s,\lfloor
x/b^{k_s-1}\rfloor)<lb\}=\min\{D_2(x):D_1(x)=s,\ x\geq l\}$$ On
the left of this value we transfer $k-1$ digits (can be
degenerated), on the right we transfer $k$ digits.

From $l_s(b-1)$ ellipses, $\frac{X_s-l}{b^{k_s-1}}$ are on the
left, $\frac{lb-X_s}{b^{k_s}}$ are on the right of $X_s$:
$$l_s(b-1)b^{k_s}=(X_s-l)b+lb-X_s=(b-1)X_s$$
We got exact formula: \be l\leq X_s=l_sb^{k_s}=lq_sb^{k_s}=lb^{c_s}<lb\ee

Intuitively the position of this boundary corresponds to the
inequality
$$b^{-k_s}\leq q_s < b^{-k_s+1}.$$
}
\item{While the situation on the top of the picture (before coding step) was fully
determined by $l$ and $q_s$, the distribution on the bottom
(after) has full freedom: is made by choosing the distributing
function.\\
This time we have the boundary value: $C(\lceil \frac
{l}{b^{k_s-1}}\rceil)\approx \frac{l}{b^{k_s-1}q_s}$.}
\end{itemize}

Finally the change of state after one step of stream coding
$\overline{C}_s: I\to I$ is:
\be \label{strc}\overline{C}_s(x):=\left\{
\begin{array}{ll}
C\left(s,\lfloor \frac{x}{b^{k_s-1}}\rfloor\right) &\textrm{for }x<X_s \\
C(s,\lfloor \frac {x}{b^{k_s}}\rfloor)\ &\textrm{for }x\geq X_s
\end{array} \right.=C(s,\lfloor x/b^{k_s-[x<X_s]}\rfloor)\approx \frac{x}{q_s b^{k_s-[x<X_s]}}\ee
we will use notation $[x<X_s]:=\left\{\begin{array}{ll} 1
&\textrm{for }x<X_s \\ 0 &\textrm{for }x\geq X_s
\end{array} \right.$.

\section{Asymmetric Numeral Systems(ANS)} In the general case:
encoding a sequence of symbols with probability distribution
$0<q_0,..,q_{n-1}<1$ for some $n>2$, we could divide the selection of
symbol into a few binary choices and just use ABS. In this section
we will see that we can also encode such symbols straightforward.
Unfortunately I couldn't find practical explicit formulas for
$n>2$, but we can calculate coding/decoding functions while the
initialization, making processing of the data stream extremely
fast. The problem is that we rather cannot table all possible
probability distributions - we have to initialize for a few of them and
eventually reinitialize sometimes.\\

This time we fix the range we are working on ($I=\{l,..,bl-1\}$),
so in fact we are interested at stream coding/decoding functions
only on this set. They are determined by distribution of symbols:
$(b-1)l_s$ appearances of symbol $s$. This way we are approximating
the probabilities. As it was already said - we will assume: $q_s=\frac{l_s}{l}$.
The exact probability will be denoted from now $q'_s$. So we have $|q_s-q'_s|\approx \frac{1}{2l}$.
\subsection{Precise coder}
We will now construct precise coder in similar way as for the binary case.\\

Denote $N_s:=\{\frac{i}{q_s}:\ i\in \mathbb{N}^+\}$.

They looks to be a good approximation of positions of symbols in the
distributing function. We have only to move them into some
positions of natural numbers. Intuition suggests that to choose
symbols, we should take succeedingly the smallest element which
hasn't been chosen yet from these sets.

Observe that $\#(N_s\cap [0,x])=\lfloor xq_s\rfloor$, but $\sum_s
\lfloor xq_s\rfloor\leq x$. So if we would use just proposed
algorithm, while choosing a symbol for given $x$, at least
$\lfloor xq_s\rfloor$ appearances of each symbol have already
appeared: $\forall_s\ x_s\geq \lfloor xq_s\rfloor$.

For $x$ being a natural multiplicity of $l$ we get equalities instead. To
generally bound $x_s$ from above, observe that because the
fractional parts of $xq_s$ sums to a natural number, we have
$\sum_s \lfloor xq_s\rfloor\geq x-n+1$.

Finally, because $\sum_s x_s=x$, we get: \be \label{prece}\lfloor
xq_s\rfloor\leq x_s\leq \lfloor xq_s\rfloor+n-1\quad\Rightarrow
x_s-xq_s\in (-1,n-1] \ee Numerical simulations suggest that there
are probability distributions for which we cannot improve this
pessimistic evaluation, but in practice $|x_s-xq_s|$ is usually
smaller than 1.\\

To implement this algorithm, in each step we have to find the
smallest of $n$ numbers. Assume we have implemented some priority
queue, for example using a heap. Besides initialization it has two
instructions: \verb"put("$(y,s)$\verb")" inserts $(y,s)$ pair into
the queue, \verb"getmin" removes and returns pair which is the
smallest with $(y,s)\leq(y',s')\Leftrightarrow y\leq y'$
relation.\\

\textbf{Precise initialization}:

\noindent \verb"For "$s=0$\verb" to " $n-1$ \verb" do {put("
$(1/q_s,s)$ \verb"); "$x_s=l_s$\verb"};"\\
\verb"For x"$=l$\verb" to "$bl-1$\verb" do"\\
\verb"  {"$(y,s)$\verb"=getmin; put("$(y+1/q_s,s)$\verb");"\\
\verb"   D[x]=(s,"$x_s$\verb") " or \verb" C[s,"$x_s$\verb"]=x"\\
\verb"   "$x_s$\verb"++}"

\subsection{Selfcorrecting diffusion (ScD)}

We will focus now on a bit less precise, but faster statistical
initialization method: fill the table of size $(b-1)l$ with proper
number of appearances of symbols and for succeeding $x$ take
symbol of random number from this table, reducing the table. So on
the beginning it will behave like a diffusion, but it will correct
itself while approaching the end.

 Another advantage of this approach is that after fixing $(l_s)$, we still have huge
(exponential in $\#I$) number of possible coding functions - we
can choose one using some key, additionally encrypting the data.\\

\textbf{Initialization}: \\
\verb"m=(b-1)l; symbols"
=$(\overbrace{0,0,..,0}^{(b-1)l_0},\overbrace{1,1,..,1}^{(b-1)l_1},..,
\overbrace{n-1,..,n-1}^{(b-1)l_{n-1}})$;\\
\verb"For "$s=0$\verb" to " $n-1$ \verb" do "$x_s=l_s$\verb";"\\
\verb"For x"$=l$\verb" to "$bl-1$\verb" do"\\
\verb"  {i=random natural number from 1 to m;"\\
\verb"   s=symbols[i]; symbols[i]=symbols[m]; m--;"\\
\verb"   D[x]=(s,"$x_s$\verb") " or \verb" C[s,"$x_s$\verb"]=x"\\
\verb"   "$x_s$\verb"++}"\\

Where we can use practically any deterministic pseudorandom number
generator, like Mersenne Twister(\cite{mer}) and use eventual key for its initialization.\\

It will be precise on the beginning and the end but generally
impreciseness will be larger. We will analyze it now. While
selecting some symbol $s$, we can divide symbols into two groups:
this symbol and the rest. So we can restrict to simplified model:\\

 \textbf{Model:} We have $N$ distinguishable numbers: $L$ copies of '1' and $N-L$ copies of '0'.
  What is the probability that if we choose $M$ of them, there
  will be $K$ of '1'?\\

$K$ copies in $M$ symbols can be distributed in ${M\choose K}$
ways. After choosing one, its copies of '1' are distributed in
$L(L-1)..(L-K+1)$ ways, of '0' in $(N-L)..(N-L-(M-K)+1)$ ways. The
number of all such sequences is $N(N-1)..(N-M+1)$, so the
probability we are looking for is: \
$$P_{N,M,L}(K)={M\choose
K}\frac{L!}{(L-K)!}\frac{(N-L)!}{(N-L-M+K)!}\frac{(N-M)!}{N!}={M\choose
K}{{N-M}\choose {L-K}}/{N\choose L}$$

For this derivation denote the expected value $q:=\frac{L}{N}$.

This probability distribution should be gaussian like with maximum
in $\frac{K}{M}\approx q$. To approximate it's width, we can use
Newton's symbol approximation from the introduction:
$$\log_2(P_{N,M,L}(K))\approx
Mh\left(\frac{K}{M}\right)+(N-M)h\left(\frac{L-K}{N-M}\right)-
Nh(q)$$

Because we are interested only in some approximation of width of
the gaussian, we have omitted terms with square root - they
correspond mainly to probability normalization. This formula
has the only maximum in $K=Mq$ as expected. Expanding around this
point up to second Taylor's term, we get

\be P_{N,M,L}(K)\approx
\exp\left(-\frac{1}{2Mq\tilde{q}}\frac{N}{N-M}\left(\frac{K}{M}-q\right)^2\right)\ee
We get mean derivative $\sigma=\sqrt{Mq\tilde{q}(1-M/N)}$.

This result agrees well with exact numerical calculations. Observe that
without $(1-M/N)$ term, it would be just the formula from central
limit theorem for the binomial distribution ($P('1')=q$).

So as expected: for small $M$ we have diffusion like behavior, but
this term makes that with $M\to N$ we approach the expected
value.\\

Returning to the algorithm, $N=(b-1)l,\ M=x-l,\ L=(b-1)l_s,\
K=x_s-l_s$:
$$x_s-l_s\approx
(x-l)q_s\pm\sqrt{(x-l)q_s\tilde{q_s}\left(1-\frac{x-l}{(b-1)l}\right)}$$
\be
x-\frac{x_s}{q_s}\approx\pm\sqrt{\frac{\tilde{q_s}}{(b-1)l_s}(x-l)(bl-x)}\label{iscd}\ee
The mean derivative is a square root of parabola with zeros in $l$
and $bl$ as expected. The maximum of this parabola will be
$\sqrt{\frac{\tilde{q}_s(b-1)}{4q_s}l}$ for $x=l(b+1)/2$. It's the largest
expected impreciseness - it grows with the square root of $l$.\\

Modern pseudorandom number generators can be practically unpredictable, so
the ANS initialization would be. It chooses for each $x\in I$
different random local behavior, making the state practically
unpredictable hidden random variable.

Encryption based on ANS instead of making calculation while taking
succeeding blocks as standard ciphers, makes all calculations
while initialization - processing of the data is much faster: just
using the tables. Another advantage of such preinitialized cryptosystem
is that it's more resistant to brute force attacks - while taking
a new key to try we cannot just start decoding as usual, but we
have to make whole initialization earlier, what can take as much
time as the user wanted. We will focus on such cryptosystems in section 8.

\section{Statistical analysis}
In this section we will try to understand behavior, calculate some properties of presented coders.
From construction they have some more or less random behavior and they process some more
or less random data so we can usually make only some rough evaluations which occurs to agree well with numerical simulations. \\

For a given coder, let us define function which measure it's \emph{impreciseness}:
\be \epsilon_s(x)=C(s,x)-x/q_s\ee
For precise coders usually $|\epsilon_s(x)|<1$, for ScD it can be estimated by (\ref{iscd}).
We have to connect it with the stream version (\ref{strc}): introduce $\overline{\epsilon}_s(x)$, such that
\be \overline{C}_s(x):=C(s,\lfloor x/b^{k_s-[x<X_s]}\rfloor)=\frac{x}{q_s b^{k_s-[x<X_s]}}+\overline{\epsilon}_s(x)\ee
$$ \overline{\epsilon}_s(x):=\left\{
\begin{array}{lll}
\overline{C}_s(x)-\frac{x}{q_s b^{k_s-1}}&=
\epsilon_s\left(\lfloor \frac{x}{b^{k_s-1}}\rfloor\right)-
\frac{1}{q_s}\left( \frac{x}{b^{k_s-1}}-\lfloor\frac{x}{b^{k_s-1}}\rfloor\right) & \textrm{for }x<X_s \\
\overline{C}_s(x)-\frac{x}{q_s b^{k_s}}&=\epsilon_s(\lfloor
\frac{x}{b^{k_s}}\rfloor)-\frac{1}{q_s}\left(
\frac{x}{b^{k_s}}-\lfloor\frac{x}{b^{k_s}}\rfloor\right) &
\textrm{for }x\geq X_s
\end{array} \right. $$
\be\overline{\epsilon}_s(x)=\epsilon_s\left(\left\lfloor \frac{x}{b^{k_s-[x<X_s]}} \right\rfloor\right)-
\frac{1}{q_s}\left\{ \frac{x}{b^{k_s-[x<X_s]}}\right\}.\ee

These equation suggest to change variable as previously: \be y:=\log_b(x)-\log_b(l)\in [0,1],\qquad
\tilde{I}:=\log_b(I)-\log_b(l)\subset [0,1],\qquad x=lb^y\ee Now
our stream coding function will be $\tilde{C}_s:\tilde{I}_s\to
\tilde{I}_s$ with $Y_s:=\log_b(X_s)-\log_b(l)$.\\
Observe that this approximated equation can be thought as
$\tilde{C}_s(y)\approx \{y-c_s\}$.

Introduce $\tilde{\epsilon}_s(y)$ analogously as before: \be
\tilde{C}_s(y)=:\left\{\begin{array}{ll}
y-c_s+1+\tilde{\epsilon}_s(y) & \quad \textrm{for}\ y<
Y_s\\y-c_s+\tilde{\epsilon}_s(y) & \quad\textrm{for}\ y\geq Y_s
\end{array}\right.\ee
Let us connect $\tilde{\epsilon}_s(y)$ with
$\overline{\epsilon}_s(y)$ and $\epsilon_s(y)$. For  $y\geq Y_s$:
\beas\tilde{\epsilon}_s(y)&=&\tilde{C}_s(y)-y+c_s=
\log_b\left(\overline{C}_s(lb^y)\right)-\log_b{l}-y+c_s=\\&=&
\log_b\left(\frac{lb^y}{q_s
b^{k_s}}+\overline{\epsilon}_s(lb^y)\right)-\log_b{l}-y+c_s\approx\\
&\approx& \log_b\left(\frac{lb^y}{q_s b^{k_s}}\right)+\frac{q_s
b^{k_s}}{lb^y\ln(b)}\overline{\epsilon}_s(lb^y)-\log_b{l}-y+c_s=
\frac{b^{c_s}}{l\ln(b)}\frac{1}{b^y}\overline{\epsilon}_s(lb^y)\eeas
where we have used the first Taylor expansion of logarithm.\\

Making similar calculation for $y<Y_s$ case, we finally get
($lb^y=x$): \be
\tilde{\epsilon}_s(y)\approx\left\{\begin{array}{ll}
\frac{b^{c_s-1}}{\ln(b)}\frac{1}{lb^y}
\left(\epsilon_s(\frac{lb^y}{b^{k_s-1}})-\frac{1}{q_s}\{\frac{lb^y}{b^{k_s-1}}\}\right)

& \textrm{for}\ y<
Y_s\\\frac{b^{c_s}}{\ln(b)}\frac{1}{lb^y}\left(\epsilon_s(\frac{lb^y}{b^{k_s}})
-\frac{1}{q_s}\{\frac{lb^y}{b^{k_s}}\}\right)& \textrm{for}\ y\geq
Y_s
\end{array}\right.\ee

\subsection{Probability distribution of the states}
We can now consider probability distribution among states our
stream coder/decoder should asymptotically obtain while processing
long stream of symbols/digits.\\

While processing some data, the state changes in some very complicated and randomly looking way. Let's remind its three sources:
\begin{itemize}
  \item Asymmetry (the strongest) - different symbols have usually different probability and so changes the state in completely different way. This choice of symbol/behaviour depends on local symbol distribution, which looks also randomly. Analogously while decoding, starting from different state, transferred bits denotes completely different behavior,
  \item Uniform covering - usually $c_s=\{\log_b(q_s)\}$ are irrational, so by making \\ $y\to^\approx\{y-c_s\}$ steps, intuitively we should cover $[0,1)$ range uniformly,
  \item Diffusion - $C(s,x)$ is near, but not exactly $x/q_s$ ($\epsilon\neq 0$), so we have some additional, randomly looking motion around the expected state from two previous points.
\end{itemize}
These points strongly suggest that the state practically behaves as random variable. So for example starting from any state, we should be able to reach any other. Unfortunately there can be found some pathological examples: in which all $\log_b(q_s)$ are rational numbers and we use precise initialization, so that we stay in some proper subset of $I$:
$$I=\{4,5,6,7\},\ n=2,\ l_0=l_1=2,\ \overline{C}_0(4)=\overline{C}_0(5)=4,\ \overline{C}_1(4)=\overline{C}_1(5)=5$$
I couldn't find qualitatively more complicated examples, but if it accidently happen the coder will still work as entropy coder, but with a bit different expected probability distribution of symbols - worse compression rate.\\

We can make natural \textbf{assumption(*)} for the rest of the paper that:
\beas&\textrm{For each two states }x,x'\in I\textrm{, there is a sequence of symbols } (s_1,..,s_m)\\&\textrm{ which makes that we go from }x\textrm{ to }x': \overline{C}_{s_m}(...(\overline{C}_{s_1}(x)))=x'.\eeas

Assume now that we want to use the coder with a sequence of symbols with given probability distribution $(p_s)_{s=0,..,n-1}$ such that $\forall_s\ 1>p_s>0$. So if in a given moment the coder is in state $x$, after one step with probability $p_s$ it will be in $\overline{C}_s(x)$ state. It can be imagined as Markov's process. Now the assumption(*) means that its stochastic matrix is irreducible - from Frobenius-Perron theorem we know that there is a unique limit probability distribution among states:
\be P:I\to (0,1),\ \sum_x P(x)=1: \ \ \forall_{x,y\in I}\ P(x)=\sum\{P(y)p_s:\overline{C}_s(y)=x\} \ee

To obtain a good understanding of the coding process, we should find a good general approximation of this probability distribution. The details of such process are extremely complicated, so to compete with this problem we should find as simple equations as possible - use logarithmic form $y=\log_b(x/l)$.

$\tilde{I}$ is difficult to handle subset of $[0,1]$, so to work
with probability on this set, we should use probability
distribution function: nondecreasing function
$\mathcal{D}:[0,1]\to [0,1]$, fulfilling $\mathcal{D}(0)=0,\
\mathcal{D}(1)=1$:
 \be \mathcal{D}(y):=\textrm{probability of being in state less or
equal than } y=\sum_{x=l}^{lb^y} P(x) \ee
It describes stationary distribution of coding process iff
$$ \mathcal{D}(y)=\sum_s p_s\left\{\begin{array}{ll}
\mathcal{D}(\tilde{C}_s(y))-\mathcal{D}(\tilde{C}_s(0)) & \quad
\textrm{for}\ y<
Y_s\\\left(\mathcal{D}(1)-\mathcal{D}(\tilde{C}_s(0))\right)+
\left(\mathcal{D}(\tilde{C}_s(y))-\mathcal{D}(0)\right) &
\quad\textrm{for}\ y\geq Y_s
\end{array}\right.$$
$$ \mathcal{D}(y)=\sum_s p_s\left\{\begin{array}{ll}
\mathcal{D}(y-c_s+1+\tilde{\epsilon}_s(y))-\mathcal{D}(0-c_s+1+\tilde{\epsilon}_s(0))
& \quad \textrm{for}\ y<
Y_s\\\mathcal{D}(y-c_s+\tilde{\epsilon}_s(y))-\mathcal{D}(0-c_s+1+\tilde{\epsilon}_s(0))+1
& \quad\textrm{for}\ y\geq Y_s
\end{array}\right.$$
We see that for $\tilde{\epsilon}=0$, the unique solution is $\mathcal{D}(y)=y$ for $y\in[0,1]$. It's idealized solution - in practice we have some discrete set of states, so $\mathcal{D}$ cannot even be continuous. $\tilde{\epsilon}$ is some very small randomly behaving function having different signs and it is somehow averaged in above equations, so intuitively $\mathcal{D}$ should be near this idealized solution. Unfortunately I wasn't able to prove it, but numerical simulations shows that this correction is in practice much smaller than $\tilde{\epsilon}$.

If we return to the original states, this approximation says that
$$P(x\leq x')\approx \log_b(x'/l).$$
Differentiating it we get that $P(x)$ is approximately proportional to $1/x$. We will use it for further calculations.

To work with $1/x$ sequences we can use well known harmonic numbers:
\be \mathcal{H}(n):=\sum_{i=1}^n \frac{1}{i}=
\gamma+\ln(n)+\frac{1}{2}n^{-1}-\frac{1}{12}n^{-2}+\frac{1}{120}n^{-4}+O(n^{-6})\ee
where $\gamma= 0.5772156649..$.
Using this formula we can easily find the normalization coefficient $\mathcal{N}$:
$$\frac{1}{\mathcal{N}}=\sum_{x=l}^{bl-1}\frac{1}{x}=\mathcal{H}(bl-1)-\mathcal{H}(l-1)\approx\ln(b)$$
For the rest of the paper we will use
\be P(x)\approx\frac{\mathcal{N}}{x}\ee
approximation. Now we can for example calculate the probability that while encoding symbol $s$ we will transfer $k_s-1$ digits:
\be P(x<X_s)\approx \mathcal{N} (\mathcal{H}(X_s-1)-\mathcal{H}(l-1))\approx \frac{1}{\ln(b)}\ln(b^{k_s}q_s)=c_s \ee
We can also define the expected value of some functions while coding/decoding process:
\be \langle f(x)\rangle=\sum_{x\in I} P(x)f(x) \approx \frac{1}{\ln(b)}\sum_{x\in I} \frac{f(x)}{x}\ee
Numerical simulations shows that they are usually very good approximations.

\subsection{Evaluation of the compression rate}
Using constructed coders we can get as near Shannon's entropy as we need. In this subsection we will evaluate this distance. It is very sensitive to parameters, so the evaluations will be very rough - only to find general dependence on the main parameters.\\

Having probability distribution of the states, we can now use (\ref{acc}) formula
\be\Delta H\approx \left\langle \frac{1}{\ln(4)} \sum_s \frac{q_s}{x^2} \left(\overline{\epsilon}_s(x)\right)^2\right\rangle\ee
Impreciseness of our encoder is more or less random and we can only estimate its expected values, so for this estimation we can threat $\frac{q_s}{x^2}$ and $\left(\overline{\epsilon}_s(x)\right)^2$ as independent variables. It would also allow to separate compression rate losses into which comes from $l, b$ parameters only and caused by impreciseness of the coder.
\be\left\langle\frac{1}{\ln(4)}\sum_s \frac{q_s}{x^2}\right\rangle\approx
\frac{\mathcal{N}}{\ln(4)}\sum_{s,x}\frac{1}{x}\frac{q_s}{x^2}\approx
\frac{\mathcal{N}}{\ln(4)}\sum_s q_s \int_l^{lb}x^{-3}dx\approx\frac{1}{l^2}\frac{b^2-1}{b^2}\frac{1}{\ln(b)\ln(4)}\ee
For the precise initialization $\left\langle\sum_s q_s (\overline{\epsilon}_s(x))^2\right\rangle$ intuitively shouldn't depend strongly on $l,\ b$ parameters, but rather on $n$ and probability distribution. Pessimistically using (\ref{prece}) we can bound it from above by $n^2$, but in practice it's usually smaller than $n$.\\

Let's focus on ScD initialization now. The term with fractional part of $\overline{\epsilon}$ is much smaller than the main source of imperfection, so we can omit it.
\beas &\left\langle\sum_s q_s (\overline{\epsilon}_s(x))^2\right\rangle\approx
\left\langle \sum_s \frac{\tilde{q_s}}{(b-1)l}\left(\frac{x}{q_sb^{k_s-[x<X_s]}}-l\right)
\left(bl-\frac{x}{q_sb^{k_s-[x<X_s]}}\right)\right\rangle\approx\\
&\approx \mathcal{N}\sum_s\frac{\tilde{q_s}}{(b-1)l}\int_l^{bl}
\left(\frac{1}{q_sb^{k_s-[x<X_s]}}-\frac{l}{x}\right)
\left(bl-\frac{x}{q_sb^{k_s-[x<X_s]}}\right)dx=\\
&=\mathcal{N}\sum_s\frac{\tilde{q_s}}{(b-1)l}
l^2\left(\frac{b^2-1}{2}+(b-1)\ln(l)-b\ln(b)\right)\approx l(n-1)
\left(\frac{b+1}{2\ln(b)}+\log_b(l)-\frac{b}{b-1}\right)
\eeas
Usually the largest is the term with $\log_b(l)$, so finally
\be \Delta H\approx \frac{\log_b(l)}{l}\frac{b^2-1}{b^2}\frac{n-1}{\ln(4)\ln(b)} \ee
Comparing to numerical simulations these estimations are very pessimistic: we get many times (like 10-100) smaller value, but general behavior $\log(l)/l$ looks to be fulfilled.\\

To summarize: in practice we rarely require that the coder is worse than optimal than e.g. 1/1000 which can be get using $l/n$ being usually below 100 for ScD initialization. Eventually we can divide $I$ into subranges initialized separately to improve preciseness.
\subsection{Probability distribution of digits and symbols}
The fact that smaller number of states are more probable unfortunately makes that produced sequences aren't exactly uniform uncorrelated sequences, what would be expected for example if we would like to use ANS in cryptography. We will analyze it briefly now and in the next section will be shown how to correct it.\\

First of all let us assume that we are coding some sequence of symbols to produce sequence of digits. Look at fig. \ref{stran}. The last transferred digit says in which subrange of states indistinguishable after bit transfer we are. So the the fact that $P(x)$ is generally decreasing, makes that it's a bit more probable that this last transferred digit is $0$. Let's estimate this probability to see how it depends on parameters.

Using $\overline{\mathcal{D}}(x):=\mathcal{D}(\log_b(x/l))\approx \log_b(x)-\log_b(l)$, we get the probability that this last (while coding)/first (while decoding) digit is $0$:
\beas &\sum_{i=0}^{(X_s-l)/b^{k_s-1}-1}\ \overline{\mathcal{D}}(l+ib^{k_s-1}+b^{k_s-2}-1)-
\overline{\mathcal{D}}(l+ib^{k_s-1}-1)\approx\\&\approx
\sum_{i=0}^{(X_s-l)/b^{k_s-1}-1}
\frac{b^{k_s-2}}{(l+ib^{k_s-1})\ln(b)}=
\frac{1}{b\ln(b)}\sum_{i=0}^{(X_s-l)/b^{k_s-1}-1}\frac{1}{i+l/b^{k_s-1}}=\\&=
\frac{1}{b\ln(b)}\left(\mathcal{H}(X_s/b^{k_s-1}-1)-\mathcal{H}(l/b^{k_s-1}-1)\right)
\approx\frac{1}{b}\log_b\left(\frac{X_s/b^{k_s-1}-1}{l/b^{k_s-1}-1}\right)=\\
&=\frac{1}{b}\log_b\left(\frac{q_sb^{k_s}-b^{k_s-1}/l}{1-b^{k_s-1}/l}\right)\approx
\frac{1}{b}\log_b\left(q_sb^{k_s}+\frac{b^{k_s-1}}{l}(q_sb^{k_s}-1)\right)\approx
\frac{c_s}{b}+\frac{1}{lq_sb^2\ln(b)}(q_sb^{k_s}-1)\eeas
where we've used $\overline{\mathcal{D}}(x+h)-\overline{\mathcal{D}}(x)\approx h\overline{\mathcal{D}}\,'(x)$ and the simplest approximation for harmonic numbers. We could get constant a few times smaller if we would take better approximation of harmonic numbers and the derivative in the middle of the range. If we are interested only in general parameters dependency, presented approximation is good enough.

In the second range probability distribution of states decreases
slower but ranges are larger. Analogous calculation gives
$\frac{1-c_s}{b}+\frac{1}{lq_sb^2\ln(b)}(b-q_sb^{k_s})$.

If we sum these values, we get that while encoding symbol $s$,
probability that the last digit while bit transfer will
be $0$ is $\frac{1}{b}+\frac{b-1}{lq_sb^2\ln(b)}$.

If we average obtained correction over all possible symbols, we
get that probability is larger than uniform digit distribution by approximately
\be \frac{b-1}{b^2\ln{b}}\frac{n}{l} \ee
In fact this value is a few times smaller and in practice we can use large $l$ like $10^5-10^6$ to make tables fit in cache memory, so this effect can be extremely weak. While estimating, probability uncertainty decreases with the square root of the number of events, so even observing this effect would require analysis of gigabytes of output. Retrieving some useful information like probability distribution of length of blocks would require much more data. For succeeding digits and correlations this effect will be accordingly smaller. We will see in the next section how to eventually reduce it as many orders of magnitude as we want.\\

Now let us focus on the opposite situation - we have some sequence of digits and we want to encode them into symbols of given probability distribution. This time states are not gathered into subranges as previously, but distributed randomly and more or less uniformly, so the differences should be much smaller. But if we need to more precisely evaluate their probability distribution than $l_s/l$, we can for example use our approximation of state probability distribution, so the probability that we will produce symbol $s$ is approximately:
\be \mathcal{N}\sum\left\{\frac{1}{x}:x\in I,\ D_1(x)=s\right\} \ee
This formula also says more precisely what probability distribution of symbols is encoded closest to the Shannon's entropy. Using it we could also modify coding/decoding functions to make better approximation of expected probability distribution of symbols using the same $l$. Shifting some appearances of symbol left(right) increases(decreases) its probability a bit.

\section{Practical remarks and modifications}
This section contains practical remarks for implementation of presented coders and some additional modifications which can improve some of their properties for cryptography and error correction purposes.

\subsection{Data compression}
Data compression programs are generally constructed in two ways:
\begin{itemize}
  \item We use constant probability distribution of symbols. It could be generally known for given type of data or estimated by statistical analysis of the file. In the second case it has to be stored in the compressed file, or
  \item The used probability distribution is dynamically estimated while encoding the file, so that while decoding we can restore these estimations using already decoded symbols. This approach is a bit slower, but we don't need to store probability distribution tables, we process the file only once and we can get good compression rates with files in which probability distribution of symbols varies locally.
\end{itemize}
ANS is perfect for the first case: using a table smaller than 100kB we can get a very precise coder which encodes about 8bits for each use of the table. It has two problems:
\begin{itemize}
  \item For each probability distribution we have to make separate initialization. We could also store tables some number of them. Observe that while changing the coder, if $b$ and $l$ are the same, we can just use the same state.
  \item Decoding and encoding are made in opposite direction - we get different sequences for estimations. To solve this problem we should process the file twice: first make the whole prediction process from the beginning to the end, then encode it in backward order. Now we can make decompression straightforward.

      In Matt Mahoney's implementations (fpaqa, fpaqc in \cite{mah}) the data is divided into compressed separately segments, for which we store $q$ from the prediction process.
\end{itemize}
For ABS situation is a bit different - we have relatively quick to calculate mathematical formulas and much smaller space of probability distributions, but we can encode only one binary choice per step. We have generally two options:
\begin{itemize}
\item Calculate formulas for every symbol while processing data -
  it is much more precise and because of it we can use large $b$ to transfer a few bits at once, but it can be a bit slower (fpaqc), or
\item Store the tables for many possible $q$ in memory - it has smaller precision, needs memory and time for initialization, but should be faster and we have large freedom of
choosing coding/decoding functions (fpaqa).
\end{itemize}
\subsection{Bit transfer and storing the tables}
For ABS using the formulas we can use large $b$, but in other cases we should rather use $b=2$. For ANS it means doing bit transfer many times in each step - this quick operation may became essential for the transfer rate of the coder. Intuition suggests that we should be able to join them into one operation per step: for example use AND with proper mask to get the bits and make corresponding bit shift right of the state.

It looks like the first problem is the order of these bits - that coding and decoding use them in reverse directions. But in fact in each step we know how many bits we should transfer and so we can just use the same direction for coding and decoding.

The larger problem is to determine this number of digits to transfer: $k_s-[x<X_s]$. It  requires the comparison and usage of small tables in which on different bits is encoded $k_s,X_s$ and maybe the mask. We could also store this information in the coding/decoding tables.\\

Let's think how to store the tables to find a compromise between memory needs and speed.

Coding tables require for given symbol $(b-1)l_s$ values from $(b-1)l$ possibilities. Usually $l_s$ isn't constant, so to optimize it for memory requirement we can encode it in one table of length $(b-1)l$: store $C(s,x)$ as \verb"C[begining[s]+x]+"$l$ where \verb"beginning[s]":=$(b-1)\sum_{s'<s}l_{s'}-l_s$. On the second side of memory/speed compromise is storing the whole $\overline{C}$. On some bits of values of this table we can store the number of transferred digits or even their sequence.

The situation with decoding tables is simpler: we can use single table of length $(b-1)l$ and store $s$ and the number of new state on it's different bits. We could also encode there the number of digits to transfer or even their sequence.\\

All these ideas require additional memory or time for using small tables. The best would be if while initialization we would generate low level code separate for each symbol - with specific $X_s$, transfers and bit masks. They can be stored such that choosing the behavior for $s$ is just a jump some multiplicity of $s$ positions.

\subsection{The initial state}
Stream coding/decoding requires choosing the initial state. The final state of one process has to be stored in the file to be able to reverse it. As it was previously mentioned - while changing coding tables, if $l$, $b$ remains the same, we don't have to change the state.

The initial state can be freely chosen - as a fixed number or randomly. We don't have to store intermediate states when we change the coding tables, but we have to store the final state. This state will be initial while decoding.

The problem could be that we are wasting a few bits in this way. Usually it should be insignificant, but for example when we want to encode separately a huge number of small files, such bits could be essential.

We can improve it by encoding some information in this initial state of the coder. We can do it for example by using a few steps of coding without bit transfer, starting from $x=0$ state. We can always do it using binary choices (ABS). Eventually we could use ANS, but it would require creating tables for additional ranges.
\subsection{Removing correlations}
In the previous section we have seen that the probability distribution of produced bits (digits) isn't perfectly uniform. It's very small effect and for correlations it would be even much smaller, but it could be significant if we would like for example use it as pseudorandom number generator. We could use some additional layer of encryption to remove correlations, but we can also do it in simpler and faster way.\\

The first idea to equilibrate probability distribution of digits is to negate (NOT) transferred digits for every second processed symbol - e.g. in steps of even number. In this way we would make that 0 and 1 are equally probable, but there would remain some correlations - '00', '11' would be a bit more probable than '01', '10'. If in one block of transferred bits we would have '0', it's a bit more probable that a few bits further (in the next block), we will have '1'.

This idea can be thought as making XOR with '00000...' and '11111...' cyclically. We can improve it by using some longer, randomly looking sequence of numbers in $\{0,..,\max_s b^{k_s}\}$ range. They can be generated using some pseudorandom number generator or even chosen somehow optimally and fixed in the coder as its internal parameters. We have to be able to recreate this sequence for decoding and store the number of last position in the file.

Now in each step of coding we take succeedingly numbers from this cyclical list and before transferring it, make XOR with the element from this list. While decoding we have to use the same list, and make XOR before using obtained bits. In this way we can reduce correlations as many orders of magnitude as we need. Blocks length varies practically randomly, so knowing this list wouldn't allow to remove this transformation.
\subsection{Artificial increasing the number of states}
Usually the number of states is $(b-1)l$, but we will see in the next chapter that sometimes it's not enough. There are generally two ways to artificially increase it exponentially:
\begin{itemize}
  \item Intermediate step(s) - the base of security of ANS based cryptosystem is that the length of blocks and the state varies practically randomly. These effects are very weakened if we want for example encrypt without compression standard data - bytes with uniform probability distribution. To cope with this problem we can for example introduce intermediate step with even randomly chosen probability distribution of symbols.

      Stream coder/decoder in one step changes a block of bits into a symbol or oppositely. We can combine such steps: decoder changes a block of bits into a symbol of given probability distribution and immediately encoder changes it into a new block of bits. Encoder and decoder have own completely separate states and modify them in opposites direction ($(y,y')\to \approx(\{y-c_s\},\{y'+c_s\})$). It looks like we move on a straight line on this twodimensional torus, but because of impreciseness, this line diffuse in the second direction and asymptotically should cover this 'torus' uniformly - the total number of states is practically the square of the original one. Surprisingly, because they use separate states, encoder and decoder can be reverses of each other.

      This approach is slower, but can be useful for cryptographic applications.
  \item Additional sequence of bits - while using ANS as error correction method, the internal state of the coder contains something like hash value of already processed message. So if it has small amount of possibilities, we can accidently get the correct value with wrong correction. The search for the proper correction requires a lot of steps, so they should be as fast as possible.

      In the previous point in each step we've changed the whole internal state of the coder - each use of a table changes one part of it, so it is relatively slow. To make it faster, we should use the table only once per step - change only part of the internal state of the coder. We could do it sequentially, but it would just separate the data into subsequences processed separately.

      The example of practical way is to expand the state of the stream coder by some cyclical table (\verb"t") of bits (eventually short bit sequences). Now coding is to make bit transfer, then switch the youngest bit of this reduced state with bit in given position in this table, increase this position cyclically and finally use the coding table. Now decoding step: use decoding table, decrease position, switch the youngest bit and make bit transfer.

\begin{tabular}{l|l}
  \textbf{Stream decoding}: & \textbf{Stream coding}(\verb"s"):\\
\verb"{(s,x)=D(x);"
&  \verb"{while(x"$\notin I_s$\verb")"\\
\verb" use s;  "
&  \verb"  {put mod(x,b) to output;x="$\lfloor$\verb"x/b"$\rfloor$\verb"}"\\
\verb" i--;switch (x AND 1)"$\leftrightarrow$\verb"t[i];" &
 \verb" switch (x AND 1)"$\leftrightarrow$\verb"t[i];i++;"\\

\verb" while(x"$\notin I$\verb")"
& \verb" x=C(s,x)"\\
\verb"   x=xb+'digit from input'" & \verb"}"\\
\verb"}"
\end{tabular}

      To make that this bit shift doesn't get us out of $I_s$, we have to enforce that $l_s$ are even. These switches increases a bit impreciseness of the coder, but if we switch only one bit, it is practically insignificant.

      This table has to be stored somehow in the output file. It's cyclical so instead of storing position, we can rotate it to make that decoding should be started with the first position.

      We see that we can in fast and simple way increase the number of internal state as much as we want. To make it faster we can represent this table of bits as one or a few large numbers.
\end{itemize}
The problem is that this large state will be required to start decoding, so we have to store it in the file. If it is used for error correction, it has to be well protected. For this purpose the initial state of the coder should be some constant of the coder, which allow to make the final verification. Eventually we could also encode some information in this initial state of the coder as previously.
\section{Cryptographic applications}
Asymmetric numeral systems were created for data compression purposes, but this simple and looking new idea of coding, has some properties which makes it very promising also for cryptography and error correction purposes. It can even fulfills all these purposes simultaneously.

\subsection{Pseudorandom number generator and hashing function}
We have seen that we can think about the state of the coder as some hidden random variable, which chooses current behavior - state change and produced bits. As we would expected from entropy coder - the output bit sequence is nearly uniform and practically uncorrelated. Unfortunately it's not perfect, but we can use not the whole state but only some of its youngest bits, what would reduce correlations greatly. Additionally we could for example use some set of masks as in the previous section.

Pseudorandom number generators (PRNG) are initialized by so called seed state: it generates randomly looking sequences, but if we would use the same seed, obtained sequence would be also the same. To use PRNG in cryptography, it has to meet additional requirements: having a sequence generated by it, we cannot get any information about the seed or further/previous bits. In the next subsection we will see that with properly chosen parameters, we shouldn't be able even to reveal the sequence of symbols used to generate random bit sequence.

So to use ANS as pseudorandom number generator we have to choose some coding function, for example initialized using the seed state. Now we have to feed it with some sequence of symbols. If this sequence is periodic, after some multiplicity of this period, the state of the coder would be the same - the bit sequence would be also periodic. But this period is much longer than the period of symbol sequence: about the number of internal states of the coder times. In the previous section we have seen that this number can be easily increased as much as we want, so in practice the sequence of symbols can be taken from some very weak psedorandom generator, or even taken as some fixed periodic sequence.\\

Hashing functions change files into some short randomly looking sequence of given length. We shouldn't be able to get any information about the file from it. Additionally we shouldn't even be able to find in practice way some two files which give the same value. To fulfill these requirements, we can for example increase the number of states of the coder by using additional table of bits as previously, process the message and for example return this table as the hash value.

If we wouldn't increase the number of states, someone could find two prefixes giving the same state and switch them. We could also prevent finding two messages with the same hash value by encoding the message twice - forward and backward. For example we can decode the file into a sequence of symbols of some fixed/generated probability distribution, then change the state and encode it back into a sequence of digits. Without changing the state we would just get the same file, but any change would make that we just produce practically random sequence - we can now for example combine some youngest bits of last used states to get the hash function.

For this purpose extremely small correlations should be completely insignificant. Eventually we could easily reduce them if we need.

\subsection{Initialization for cryptosystem}
For given parameters we still have huge amount of coding functions with practically the same statistical properties, but producing a completely different encoded sequence. If we make selfcorrecting diffusion initialization using some PRNG initialized using given key, we would get practically unique coding function for this key. If we would use it to encode some information, it looks practically impossible to decode the message not knowing the key. We will now make a closer look at such approach to data encryption.\\

First of all, let us focus on the ScD initialization. It's large number of picking a random symbol from some large table. The coding table is approximately given by symbol probability distribution, but it looks practically impossible to find its precise values not knowing the key. The initialization process strongly depends on its history, which creates specific symbol distribution in the \verb"symbols" table - while knowing the key, it looks practically impossible to find $C(x,s)$ without making whole previous initialization (for smaller $x$).

So to start decoding we practically have to make whole initialization. Observe that we can enforce PNRG to require as large time to be calculated as we want, for example:

\verb"for i=1 to N do {k=random; read k random values}"\\
makes that we statistically know approximately in which position of PRNG we will be. But to find the the exact position, we just have to make all calculations.

We see that this way we can enforce some time required for initialization. Connecting it with the unpredictability of ScD initialization, we see that such cryptosystem would be extremely resistent to brute force attacks. Standard approach makes all computations while processing the file, so to check if given key is correct we can just start decrypting the file and observe if the output for example isn't a completely random sequence. In the presented approach, most of computation is made while initialization: to check if given key is correct we have to spend given time to make the initialization, for example enforced to take about 0.1s - it's a few orders of magnitude larger than in standard approach. After initialization the processing of the data uses already calculated tables - is much faster than in standard approach.\\

Now assume that someone would get the coding function - does it mean that he can retrieve the key? This function says symbols chosen in each steps, but each symbol could be chosen in many ways, so in fact he wouldn't have sequence of used random variables, but only some sets of its possible values - even using a weak PRNG it looks practically impossible to deduce the key. Eventually we could use some secure PRNG, for which it is ensured that knowing the exact sequence, we couldn't find its seed state and so the key.

This property suggests extremely powerful additional protection - use not only the key as the seed state, but also some number which can be even stored in the file. Now after every encrypted fixed number of bits (like gigabyte), we change this number, store it and use to generate new coding tables. The size of these blocks should be chosen so that it wouldn't be possible to retrieve any essential information from them. The behavior of each one is practically unrelated, so their information couldn't be connected for finding the key.

Sometimes we would like to make encryption and entropy coding in the same time. The question is - what to do with the probability distribution of symbols. There wouldn't be a problem if we would use some adaptive prediction method, but it would also require using many different coding tables. We will see that these tables should be rather large, so sometimes it might be better to use fixed probability distribution of symbols. It has to be stored in the header and so is easily accessible. We will see that such knowledge shouldn't rather make breaking the code easier, but it gives some knowledge of file content, what can be unwanted. To prevent it, this header can be encrypted separately using the same key but probably in some different way.

\subsection{Processing the data}
The coder uses the state which is hidden practically random variable. Also hidden, randomly generated local behavior of the coding function defines current behavior - how many digits to produce and to which state jump. Blocks created this way are relatively short, but they have various, practically randomly chosen length. This picture looks perfect, but unfortunately there are some weaknesses which could give some information about statistics of symbols or even coding function. They would vanish if we would use some additional layer of standard encryption, but I will try to convince that using only ANS with proper parameters and some quick and simple modifications, we can make really safe and fast encryption.

\begin{itemize}
\item First of all, as it was mentioned in the previous section - the base of the randomness of the state is that we \textbf{don't use uniform distribution of symbols} (asymmetry) and that some symbols has probability not being an integer power of $b$. These assumption is in practice automatically fulfilled when we make encryption and entropy coding in the same time, but sometimes we would like just to encrypt some more or less uniform byte sequence. The best way to cope with this problem is to use the intermediate step from the previous section - using the same PRNG choose some probability distribution of symbols and then in one step encode a byte into a symbol and immediately use it to produce output bit sequence. Alternatively if we want to make it quicker - use only one step: we can use the same PRNG to modify randomly the uniform distribution among bytes a bit and treat input sequence as sequence of such symbols. The cost is that the state doesn't change as fast as previously and that the output file is a bit larger than the input. We have also smaller amount of possible states of the coder this way. Eventually we could use so called homophonic substitution - to each symbol assign a few new ones and choose among them using some separate (hardware) random number generator, but it would increase the size of the message.
\item If we would encode the same sequence starting from the same state of the coder, we will get the same output. To prevent attacks based on such situations, we should \textbf{increase the amount of its internal states}. In the previous section were shown some ways to do it - use some correlation removing method, intermediate step or additional bit sequence.
\item As it was previously mentioned, because the probability of being in given state ($x$) is not uniform, but is decreasing ($\propto 1/x$), some produced blocks of digits are a bit more probable (with smaller digits). These differences are extremely small and because of various block length I don't see a way to use it to find given block structure or some precise information about coding function. But analyzing statistically huge amount of data, one could evaluate probability distribution of block lengths, which gives some information about probability distribution of symbols. To prevent it we can use some of presented method of \textbf{removing correlations}. We could also generate sometimes new coding function for example with the same key but with some new additional, presented number.
\item Transferred digits are the youngest digits of the state. If one would have both ciphertext and corresponding plaintext, would make a correct assumption about the internal state and blocking in given moment and knew precisely used probability distribution of symbols, he could track the history of the processing, which would reveal used coding table. Let's focus on such scenario. Knowing probability distribution of symbol, we know that $x\to^\approx \frac{x}{q_sb^{k_s-[x<X_s]}}$. If we used ScD initialization, the impreciseness of such prediction of $x$ is of $\sqrt{l}$ order. The transferred digits give precise position in the range of width $b^{k_s-[x<X_s]}$ ($1/q_s$ at average).  So if $l$ \textbf{is large enough}, \be l>q^{-2}_s\ee in presented scenario the number of possibilities the person would have to consider would grow exponentially, making such attack completely impractical. Observe that $q_s$ is at average $1/n$, so above condition tells also that $l>n^2$.

     In practice any presented method for increasing the number of internal states should also prevent such scenarios.
\item Having a lot of ciphertext and corresponding plaintext, one could try to make some statistical analysis to connect symbols with blocks. Because of various length of blocks it doesn't look practical, but to prevent such eventualities it would be expected that every symbol could produce practically all possible youngest digits of the state. Given symbol can produce $(b-1)l_s=(b-1)lq_s$ different states and in importance (shown in the encrypted file) are let say $k_s$ youngest digits ($1/q_s$ values at average), so we again get $\sqrt{l}>1/q_s$ condition. Again any other modification would also give good protection.
\item If one can use initialized coder (adaptive scenario) and has some message encrypted with it, he could try to use different inputs, slowly exposing succeeding bits of the plaintext. This is unavoidable weakness of using short block length cryptosystems. Fortunately there is simple universal protection against such rare scenarios: add a few random bytes at the beginning of the file before starting encryption or choose the initial state randomly (this time not using PRNG used for initialization). In this way while encrypting the same data, we will most probably get different output, which still can be decrypted into the same input data.
\end{itemize}
I cannot assure that this list is compete, but for this moment I cannot think of more weaknesses which could be used to break ANS based encryption. We can easily protect against all of them.

To summarize, while designing a cryptosystem base on ANS, we should:
\begin{itemize}
  \item Ensure the asymmetry - that the probability distribution of symbols is not uniform,
  \item Use $b=2$ for which state probability distribution is nearest uniform,
  \item Use large $l>n^2$ or even $\forall_s\ l>q_s^{-2}$. So to make coder faster (larger $n$), we should use correspondingly large tables,
  \item Use some correlation removing modification and eventually increase additionally the number of internal states of the coder,
  \item Eventually choose randomly the initial state of coder.
\end{itemize}

\section{Near Shannon's limit error correction method}
While compressing a file we remove some redundancy caused by statistical properties. Using forward error correction methods, we are adding some easily recognizable redundancy, which can be used to correct some errors. In standard approach we usually divide the message into short independent blocks. The problem is that it's vulnerable to pessimistic cases (large local error concentrations) - if the number of errors exceeds some boundary, we loose the whole block. In this section will be shown how to connect their redundancy to treat the whole message as one block. I will focus on using ANS for this purpose, but presented approach is more general - fig. \ref{conred} shows how to use it for any block code and a hashing function or even only a hashing function.

For simplicity let us assume the simplest channel for this paper: memoryless, symmetric. That means that there is some fixed probability ($p_b\in [0,0.5)$) that transmitted bit will be changed ($0\leftrightarrow 1$). So while transmitting $N$ bits, about $Np_b$ of them will be damaged.\\

For a channel of given statistics of errors (noise), we can say about Shannon's limit - theoretical maximal information transfer rate. Constructions used to show that this limit is achievable are completely impractical. Near this limit are Low-Density Parity-Check Codes (LDPC) (\cite{gal},\cite{mackay}), but they still they require solving NP-problem to correct. So in practice there are used codes which divide the message into independent blocks, what makes them vulnerable to pessimistic cases. For example for $p_b=0.01$, we should be able to construct a method which adds asymptotically a bit more than 0.088 bits of redundancy/transmitted bit and is able to fully repair the message. Compare it with commonly used (7,4) Hamming codes - it adds 3 bits of redundancy per 4 transmitted bits to be able to correct 1 damaged bit per such 7 bit block. It uses much more redundancy: 0.75 bits/transmitted bit, but because sometimes we have more than one error in block, we loose about 16bits/transmitted kilobyte and we don't even know about it.\\

Imagine we have some channel with known statistical model of error distribution. To transmit some undamaged message through it, we have to add some redundancy 'above' given error density. We know only statistics of errors, not when exactly they will appear - so this density of redundancy should be chosen practically constant. But the density of errors fluctuates - sometimes is locally high, sometimes low. We see that while dividing the data into independent blocks, we have to choose the density of redundancy accordingly to some pessimistic error density in such block. But in fact there usually isn't some pessimistic level - we only know that the worse case, the rarer it occurs. So in this way for most of blocks were used much more redundancy then required, but for some of them this amount is still not sufficient.

We see that to obtain a really good correction method, we should treat the message as the whole. In LDPC it is made by distributing uniformly some large amount of parity checks. Presented approach divides the message into blocks, but their redundancy is connected by the internal state of the coder, which contains something like checksum of already processed message to choose local behavior. Using these redundancy connections we can intuitively 'transfer' surpluses of unused redundancy to cope with pessimistic cases. We will see that we are able to get near Shannon's limit this way with practically linear expected time of correction.
\subsubsection{Advantage of connecting redundancy of blocks}
\begin{figure}[h]
    \centering
        \includegraphics{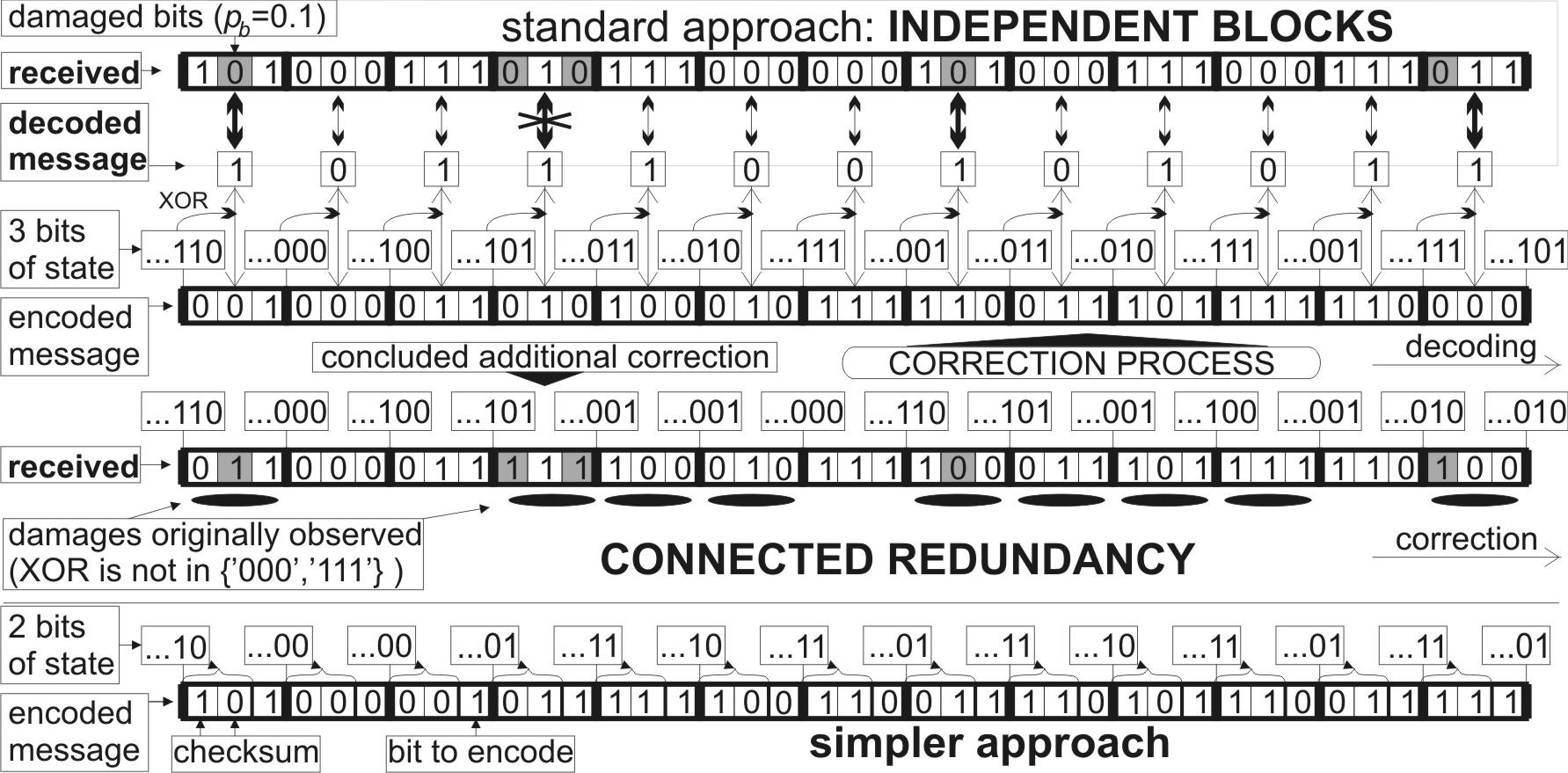}
        \caption{Two simple schemes of block codes with connected redundancy. On the top there is example of usage of standard triple modular redundancy block code - we send three copies of each bit and decode as the value with more appearances. In the middle it is shown how to modify it to connect redundancy of blocks - we use hash value of already processed encoded message stored in the state of decoder - modify the original block by making XOR with some 3 bits of this state. On the bottom there is simpler approach, which don't need a standard block code - in one block we place some 2 bits of the state of decoder (kind of checksum) and the bit to encode in the third one.}
        \label{conred}
\end{figure}

There will be now shown two approaches of connecting redundancy of blocks as in fig. \ref{conred}. In fact we will do it later in more flexible and usually faster way, but these approaches show advantages of this new correction mechanism. Analysis and methodology from further subsections apply also to these approaches.

The basic tool to connect redundancy of blocks is some hash function, which allows to deterministically assign some shorter, practically random bit sequence to already processed encoded message. In practice it is usually done by some automate which has a state containing such hash value to given position and changes this state while processing succeeding portions of the message.

Look at the middle of the figure - after using the original block code (like $1\rightarrow 111$), we make XOR with some bits of this state, containing practically random bits determined by already processed message (like XOR$(111,110)=001$). Now while decoding - if given block wasn't damaged and the state is correct, XOR of the state and the block should be a codeword ($000$ or $111$). If not - we know that there was a damage. Now as long as in each block there is at most one damaged bit, we immediately know which bit we should repair - the behavior is similar as for independent blocks. The advantage starts when there occurs more errors in one block - the state from this moment will most probably be different than expected and so for each block with some probability ($p_d=1-2/2^3=3/4$), we should observe that it's damaged - it's much more often than expected for the proper correction and so suggests where to search for additional corrections.

The bottom example shows that we don't really need to base on a block code - in some positions of the block we place some bits of the state(checksum) and bit(s) we want to encode in the rest of them. Now we immediately observe if positions with checksum were damaged. If the essential bits were damaged, we will conclude it later thanks of this new correction mechanism ($p_d$ is still 3/4). We will see that as it suggests - using only this new correction mechanism, we can already get near Shannon's limit. In the last subsection will be shown how to connect it with block codes mechanism as in the middle picture to correct simple damages immediately, additionally quickening the correction process.

\subsection{Very short introduction to error correction}
Forward error correction can be imagined that among all sequences, we choose some allowed ones - so called codewords. They have to be 'far' enough from each other, so that when we receive a damaged sequence, we should be able to uniquely determine the 'nearest' allowed one. Of course we would also want that the probability that it's really the correct sequence is large enough. In other words we divide the space of all sequences into separate subsets - kind of balls around codewords. The 'thicker' these balls are, the larger probability that we make the correction properly.

In standard approach we divide the message into blocks of fixed length, which are encoded independently. In this case we can use Hamming distance - the number of positions on which given two sequences of bits differ. For example triple modular redundancy code uses 3 bit sequence to encode 1 bit - in the space of $2^3$ possible 3 bit sequences, there are chosen $2$ codewords ($000$, $111$), which are centers of balls of Hamming radius 1. So if while transmitting given 3 bit block, at most one bit was changed, we can correct it properly. If the number of changed bits is larger than one, we get into a different ball - it is corrected in wrong way and we even don't know it.\\

Let us focus on a memoryless symmetric channel: if we received '1', with probability $1-p_b$ it was really '1' and with probability $p_b$ it had to be '0'. If we would know in which of these cases we are, we would get exactly one bit of information. To distinguish them is needed $h(p_b)=-p_b\lg(p_b)-(1-p_b)\lg(1-p_b)$ bits of information, so such 'uncertain bit' contains $1-h(p_b)$ bits of information - to transmit $N$ real bits, we have to transmit at least
\be N\frac{1}{1-h(p_b)} \ee
such 'uncertain bits' - it's so called Shannon's limit and the channel coding theorem says that theoretically we can get as near as we want to this capacity. It means that for a channel with given statistics of error, we should be able to construct an error correction method which uses a bit more than $\frac{1}{1-h(p_b)}-1=\frac{h(p_b)}{1-h(p_b)}$ bits of redundancy per bit of message and is able to completely repair the message.\\

While working on such potentially infinite blocks, the number of damaged bits tends to infinity, so we can no longer work on the Hamming distance. Now while transmitting given codeword $T$ of length $N$ bits, will be received a message $R$ with damaged bits on some more or less $Np_b$ positions. The positions of these errors can be stored as length $N$ bit sequence $E$, such that \be R=T\oplus E  \label{rtd}\ee
where $\oplus$ denotes addition modulo 2 of two bit vectors (XOR). This $E$ vector statistically should be chosen as one of ${N\choose {Np_b}}\approx 2^{Nh(p_b)}$ possibilities. From (\ref{rtd}) we see that it's also the number of possible received sequences corresponding to one codeword. If we will divide the number of all possible received sequences by this number, we can get Shannon's limit again: $2^N/2^{Nh(p_b)}=2^{N(1-h(p_b))}$.

In fact the number of damaged bits is close but rather not exactly equal $Np_b$. But if we have some large number ($N$) of independent identically distributed random variables of entropy $H$, their outcome is almost certain to be in some set of size $2^{NH}$, which all members have probability 'close to' $2^{-NH}$ - it's so called 'asymptotic equipartition' property (\cite{mackay}). This set is called \emph{typical set}, for example:
\be \left\{x\in \{0,1\}^N:\left|\frac{1}{N}\lg\left(\frac{1}{ p^{\#\{i:x_i=1\}}(1-p)^{\#\{i:x_i=0\}}}\right)-h(p)\right|<\beta\right\}.\label{typic}\ee
For all $\beta>0$ and correspondingly large $N$, such set contains almost whole probability. Subrange of typical set is asymptotically also typical, so these practically $Np$ copies of '1' should be spread more or less uniformly. \\

Shannon's coding theorem says that we can get as close to the theoretical limit as we want and we should be able to correct practically all possible typical errors. So we should look for the proper correction among typical ones with $p_b$ probability of '1'. Standard proof generates the set of codewords randomly, modify this set (remove some codewords) and show that for large $N$, with probability asymptotically going to 1, we can properly determine transmitted codeword. Unfortunately it would require to check exponentially large set of possible corrections - it rather cannot be done in practice.
\subsection{Path tracing approach}
First of all, let us focus on sketch of different but still impractical proof: using a hashing function. Such function allows to assign to each message some shorter, practically random bit sequence. Assume now that we transmit the original message of length $N$ bits through the channel and its 'a bit longer' than $Nh(p_b)$ bits hash value through some different noiseless channel. Now the receiver can check 'all typical corrections' ($2^{Nh(p_b)}$) of the received message and almost certainly only one (the proper one) will give the expected hash value. If we would like to transfer the hash value through the same noisy channel, we can analogously send additionally its hash value and so on (until it's smaller than some limit value which can be encoded in some different way). So finally we would asymptotically need at least $$N(1+h(p_b)+h^2(p_b)+...)=\frac{N}{1-h(p_b)}\quad \textrm{bits}$$
Observe that 'a bit longer' can mean that the number of hash values is larger only polynomially with $N$ than the number of typical corrections - there still almost certainly will be only one proper typical correction and we will get asymptotically exactly Shannon's capacity.

There has left to precise what does 'all typical corrections' mean. For a theoretically infinite data stream we should be able to take $\beta\to 0$ limit in (\ref{typic}). It can be achieved by intersecting sets of corrections passing verification for some sequence $\beta_i\to 0$. For a data stream of a finite length, we could take proper correction which is nearest to typicality (smallest $\beta$), but we will see that the best will be such that corrects the smallest number of bits.\\

We will now modify this method to make it practical - instead of making huge verification once, intuitively we will spread it uniformly over the whole message. Thanks of it we will be able to detect errors not only on the end of the process, but also shortly after they appear: after an error in each step we have some fixed probability ($p_d$) to detect that something was wrong. We will pay for this parameter in capacity, but when it exceeds some critical point, the number of corrections not detected by this mechanism will no longer grow exponentially. So we will no longer require that the amount of possible hash values (states of the coder) should grow exponentially - the relative cost of storing it will vanish asymptotically.

This threshold corresponds to the Shannon's capacity, but in practical (nearly linear) correction methods there appears some additional problem with large errors concentrations and we should use a bit larger capacity. We will see it in the next subsection.

\begin{figure}[h]
    \centering
        \includegraphics{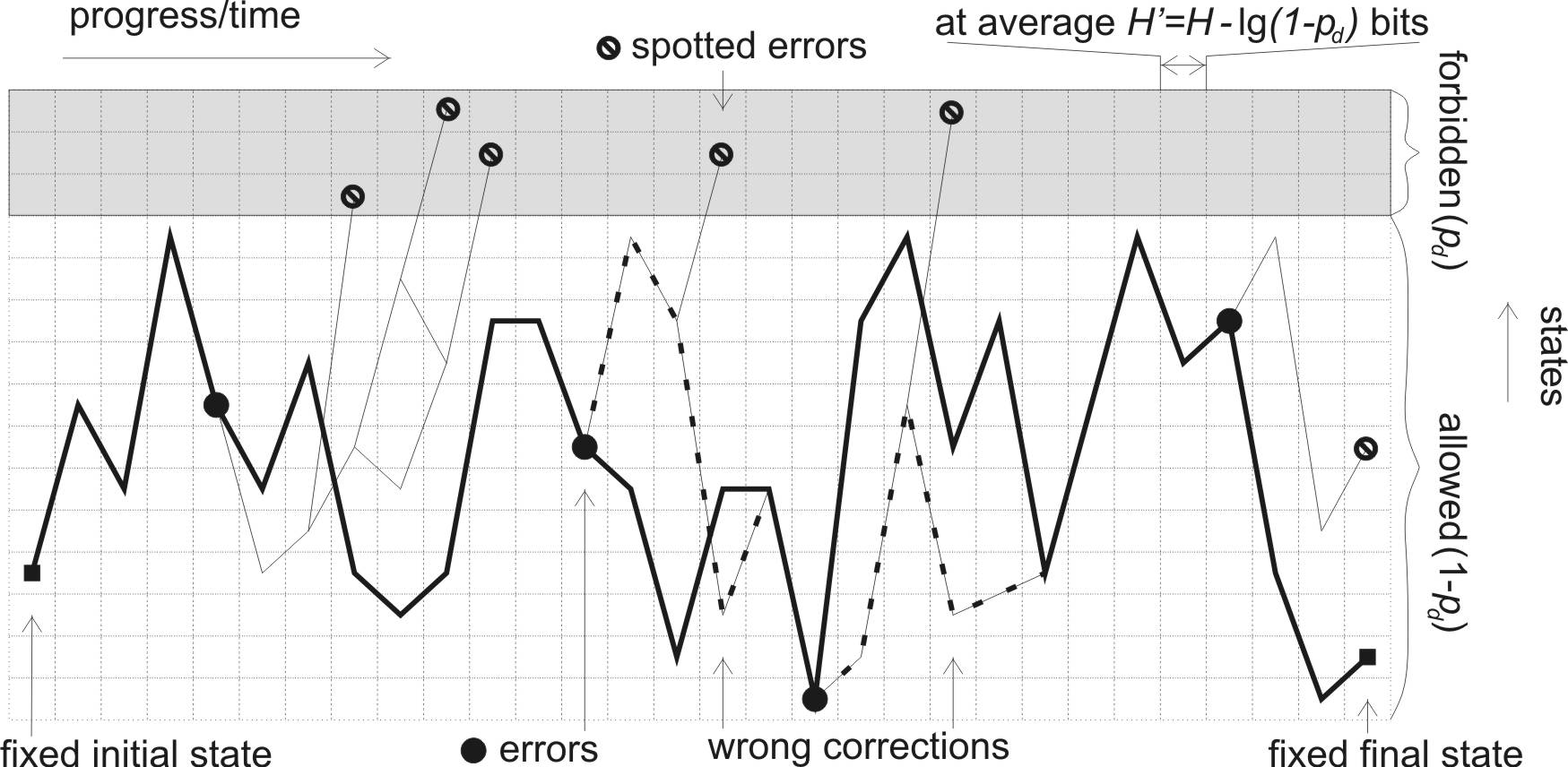}
        \caption{Schematic picture of path tracing correction algorithm. If the number of states is large enough, corrections to consider should no longer create cycles as in the figure and so create tree.}
        \label{pathtrace}
\end{figure}

The situation looks like in fig. \ref{pathtrace}: the transmitted codeword (correct path) is denoted by the thick line. We start with the fixed initial state and try to process succeeding bits. While we are on the correct path, the state changes in randomly looking way among all allowed states. After an error (we've lost the path), the state also changes in randomly looking way, but this time among all states - in each step there is some probability ($p_d$) that we will get to a forbidden state (observe that something was wrong).

The problem is that after an error, before it will be detected, the coder can accidently get into the correct state - we would go back to the correct path without a possibility to detect that we've made a wrong correction. We will see later that with proper selection of parameters, errors will appear slower than we can correct them - probability of such situation will drop asymptotically to zero.\\

We can use entropy coder with internal state for such path tracing purpose: add a forbidden symbol of probability $p_d$, marking its appearances as forbidden states and rescale correspondingly probabilities of the rest of symbols (allowed ones). We could eventually use arithmetic coder, but ANS is faster, has useful modification capabilities and is generally simpler, so I will concentrate on it.

If we want to encode a symbol sequence with $(q_s)_s$ probability distribution, we have to use correspondingly $((1-p_d)q_s)_s$ probability distribution instead. Now while encoding we use only these allowed symbols. If there wouldn't be errors, while decoding we would also use only allowed symbols, but after an error we would produce practically random sequence of symbols, so in each step we have probability $p_d$ of trying to produce the forbidden symbol and so detecting that there was an error.

To use ANS for this purpose we would rather need to use some method to increase the number of internal states of the coder to reduce probability of wrong correction. The initial state can be generally fixed for encoding process, but the final one has to be stored, probably in the header of the file. So there have to be used some separate strong error correction method for it to make that we can be sure that this initial decoder state is proper.\\

What is the cost of adding such forbidden symbol? The data sequence contains at average $H=-\sum_s q_s \lg(q_s)$ bits per symbol. After the rescaling, we will use at average
\be H':=-\sum_s q_s \lg((1-p_d)q_s)=H-\lg(1-p_d)\quad \textrm{bits/symbol}.\ee
There will be shown now intuitive argument that choosing $p_d$ as in Shannon's limit:
\be -\lg(1-p_d)\geq H\frac{h(p_b)}{1-h(p_b)} \label{sl}\ee
the possible space of hash values (states of the coder) wouldn't longer have to grow exponentially as for $p_d=0$ from the beginning of this subsection. In this case the cost of storing this (protected) hash value would vanish asymptotically. Denote \be p_d^0=1-2^{-\frac{H}{1-h(p_b)}h(p_b)} \qquad\qquad\left(=1-2^{-H'h(p_b)}\right)\ee
this threshold value. For simplicity let us assume that $b=2$.

Unfortunately there is a very subtle problem with this argument, which will be explained and precisely analyzed in the next subsection.\\

Assume we've received some message of length $N$ bits. We will use the simplest method in this moment: as before try to correct it using 'all typical corrections' and check if they pass the verification: while decoding we would use only allowed states and the final state is correct. As in the picture, such message agrees with the correct one before the first error. Then it can vary according to the noise until it reaches a forbidden state or the correct state for given point.

Let us assume that in $j>0$ steps after an error it still didn't reach a forbidden state. If we wouldn't accidently get to the allowed state, the probability of such situation is $(1-p_d)^j$. One step corresponds to at average $H$ encoded bits which correspond to at average $H-\lg(\tilde{p}_d)$ transmitted bits, so in $j$ steps we processed at average $j\left(H-\lg(\tilde{p}_d)\right)$ bits. They can freely change according to the noise, so we should check about $2^{j\left(H-\lg(\tilde{p}_d)\right)h(p_b)}$ their corrections. If we choose $p_d$ such that
\be 1\geq (1-p_d)^j 2^{j\left(H-\lg(\tilde{p}_d)\right)h(p_b)}=
\left(2^{\lg(\tilde{p}_d)+\left(H-\lg(\tilde{p}_d)\right)h(p_b)}\right)^j \label{naive}\ee
the expected number of such corrections not rejected by this mechanism will no longer grow exponentially. This threshold is exactly the Shannon's limit (\ref{sl}).

We can now intuitively estimate the probability of wrong correction scenarios as in the picture - that we can start with an error from a correct state in some point of time and after some typical noise accidently get back to some correct state. There are almost $N$ possible starting points for such scenario. If $p_d\geq p^0_d$, the expected number of corrections which errors won't be detected by the forbidden states mechanism doesn't longer grow - it usually even drops to zero with the width of such subrange ($j$). So the expected number of such scenarios can be bounded from above by $N^2$. If the number of states of the coder behaves for example like $N^3$, almost certainly only the proper correction will pass the verification.

To store protected one of $N^3$ values we need about a bit more than $3\lg(N)$ bits - while calculating channel's capacity this cost vanishes asymptotically.

\subsection{Practical correction algorithms}
Methods constructed on proofs of that we can approach Shannon's limit requires clearly exponential correction time, like checking all typical corrections. In this subsection I'll try to convince that practical (nearly linear) correction methods have to require redundancy level above some found higher limit. Then there will be presented general approach to correction - building correction trees. For basic choice of weight it requires a bit more redundancy than this new limit.
\subsubsection{Practical correction limit}
Generally if we want to find correction in practically linear time, we rather cannot work on corrections of the whole message (exponential number), but should slowly elongate them. So practical correction algorithms should use enough redundancy to ensure that expected number of corrections to be considered up to given point is finite.

The problem is that in fact we don't know the number of damaged bits, only that they appear with $p_b$ probability - that asymptotically probability distribution of the number of damaged bits is Gaussian with $\sqrt{Np_b (1-p_b)}$ standard deviation. This uncertainty vanishes asymptotically, but surprisingly has essential influence on the expected number of corrections to be considered in given moment (width of the correction tree) - rarely there are very large local error concentrations which result in infinite expected width - it essentially influence (\ref{naive}).\\

We will see it formally later, but intuitively among corrections which survived to given moment, the less bits they changed, the more probable they are. It suggests the simplest correction algorithm - moving the 'front' of the tree - for given moment remember some number ($M$) of corrections which passed verification with the smallest number of corrected bits. Now in each step try to expand all of them and take only the best $M$ of those which gave some allowed state.\\

The question is: how many of them ($M$) we should remember?\\

In other words - which in this order is the proper one? The larger $M$ is, the slower the algorithm, but also the larger probability that there is the proper correction among considered ones. If in given moment this number is not enough, we lost this proper correction. Fortunately we can observe it - from this moment the number of damages will have to be larger than expected (about $h^{-1}(-\lg(\tilde{p}_d)/H'$)) - it can suggest to go back and use locally larger $M$. Later we will do it smarter, but for this moment assume that we use always large enough various $M$.

To summarize: we have to make that the expected number of corrections which passed the verification up to given moment ($J$ bits) and is changing smaller number of bits than the proper correction is finite.

The probability distribution of the number of damaged bits is Gaussian with center in $p_bJ$ and standard deviation $\sqrt{Jp_b (1-p_b)}$. If there was in fact $pJ$ damaged bits, the number of wrong corrections with at most this number of damaged bits will be asymptotically dominated by ${J\choose{pJ}} \approx (2\pi
Jp\tilde{p})^{-1/2}2^{Jh(p)}$ and about $\tilde{p}_d^{J/H'}$ of them are expected to survive these about $J/H'$ steps.

So the expected number of those which survived is asymptotically approximately
$$\int_0^\infty (1-p_d)^{\frac{J}{H'}}\cdot\frac{1}{\sqrt{2\pi
Jp\tilde{p}}}\ 2^{Jh(p)}\cdot\frac{1}{\sqrt{2\pi
Jp_b\tilde{p}_b}}\ e^{-\frac{(pJ-p_bJ)^2}{2Jp_b(1-p_b)}}\ J dp$$
It is finite for $J\to\infty$ if
$$\frac{1}{H'}\lg(1-p_d)+h(p)-\frac{\lg(e)}{2p_b(1-p_b)}(p-p_b)^2 < 0$$
This function of $p$ has only one maximum - a bit above $p_b$ as expected. This $p$ corresponds to cases which influence most the expected number of corrections.

Finally the integral is finite if we chose $p_d > p_d^1$:
\be p_d^1:=1-2^{-H'\max_{p\in[0,1]} \left(h(p)-\frac{\lg(e)}{2p_b(1-p_b)}(p-p_b)^2\right)}
\qquad\qquad\left(\,\geq p_d^0\,\right) \ee
Using $H'=H-\lg{\tilde{p}_d}$, we get
\be \frac{H'}{H}=\frac{1}{1-\max_{p\in[0,1]} \left(h(p)-\frac{\lg(e)}{2p_b(1-p_b)}(p-p_b)^2\right)} \ee
Comparing to Shannon's limit - it's a bit larger for small $p_b$, up to twice larger for $p_b\to 0.5$. This formula can be used to choose $p_d=1-2^{H(1-H'/H)}$.
\begin{figure}[h]
    \centering
        \includegraphics{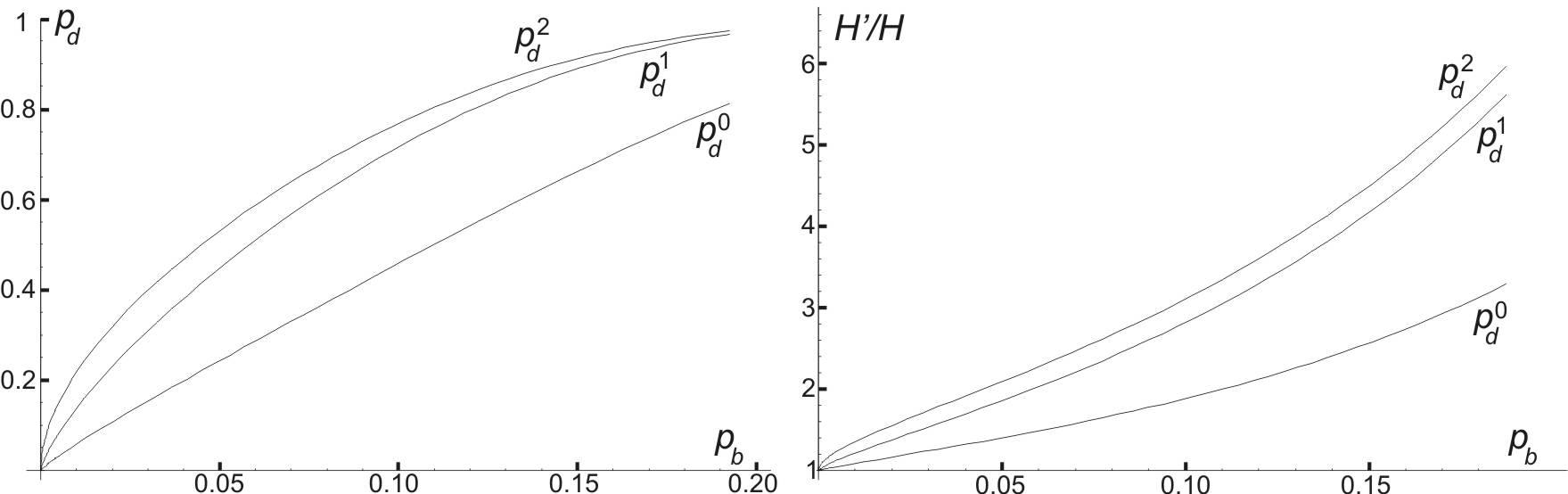}
        \caption{Comparison of Shannon's limit ($p_d^0$), limit for practical correction algorithms ($p_d^1$) and reached by the basic version of correction tree algorithm ($p_d^2$).}
        \label{limits}
\end{figure}

To summarize: if we would use $p_d \in [p_d^0,p_d^1]$, while trying to consider some best corrections up to given moment, we asymptotically would need exponential number of steps. Because we rather use polynomially large number of states of the coder, eventually found correction will be probably wrong. If we can tolerate exponential correction time, we can use exponential number of states and smaller $p_d$ getting better limit and finally Shannon's limit for $p_d=0$ as in the beginning of the previous subsection.

In practice we use finite blocks, so choosing some large $M$, with large probability it will find the proper correction (in linear time). We can verify it: it's most probably the proper correction, if the density of damages didn't drastically grow from some point. We could focus on this point and try to correct this correction. It suggests how to choose $M$: use relatively small $M$ first until error density grows much faster then it should. Now interpolate this point and try to use larger $M$ around until error density won't return to expected.

\subsubsection{Correction tree algorithms}
Previously suggested algorithm was considering some number of best corrections to given point - we are expanding the tree of possible corrections by moving its whole 'front'. We will now look for some sophisticated algorithms - which in given moment selects some best node to expand. We should get some (pseudo)random tree with many subtrees of wrong corrections growing from the core made of the proper correction. These subtrees has generally larger error concentration.

\begin{figure}[h]
    \centering
        \includegraphics{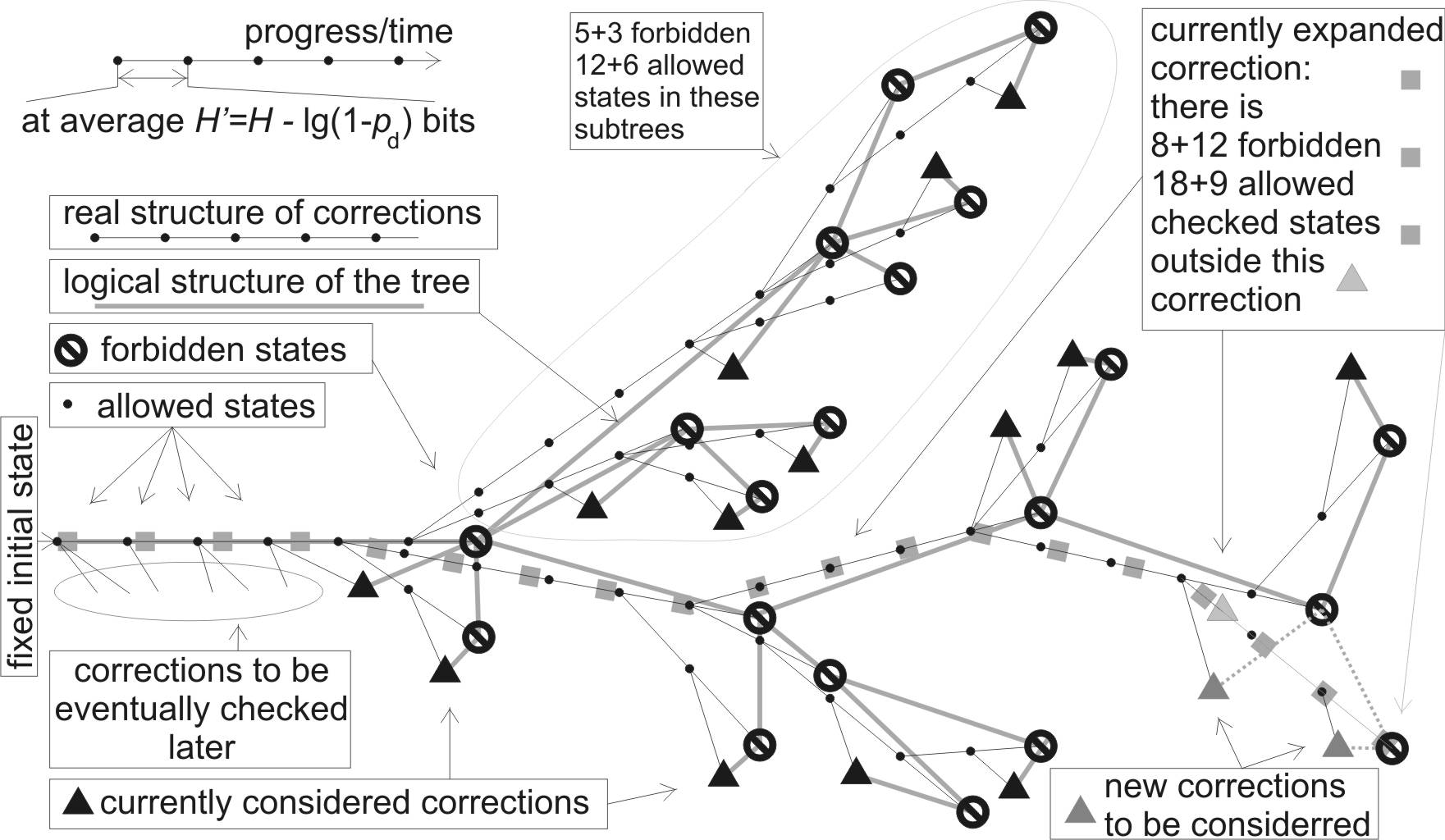}
        \caption{Correction algorithm. It will create such (pseudo) random trees. If $p_d<p_d^0$ this tree would immediately grow exponentially. If $p_d=p_d^0$ it would look to grow linearly. If $p_d^0<p_d\leq p_d^2$ and we use basic weight function, its width will be generally small, but rare high error concentrations will sum to infinite expected width. If $p_d>p_d^2$ its expected width will be finite - it can be used for potentially infinite messages. This limit can be probably improved up to $p_d^1$.}
        \label{coralg}
\end{figure}
Each dot in fig. \ref{coralg} corresponds to some allowed state of decoder. When we get to a forbidden state, we have to expand somewhere else. Edges of such tree correspond to some corrections of bits used in given step - intuitively the less bits it correct, the more probable it is - we should try it earlier.

To create logical structure of the tree, we have generally three possibilities:
\begin{enumerate}
  \item{make node for each bit - split denotes changing one bit - it's rather impractical, or}
  \item{make node for each step - split denotes correction of bit block used while the last step - good for large $p_b$, or}
  \item{make node every time forbidden state occurs (as in the figure) - we connect consecutive bit blocks as long as we can decode further without correction - good for small $p_b$.}
\end{enumerate}
In each node we have to store somehow the pointer to its father, lately used correction and the position in the message. To choose quickly the best node for given moment, we will store there also some wight and always choose node with the largest one. For each node we can easily find its most probable, not considered yet child - they are denoted as 'triangles' in the figure. In given moment we can focus on them and mark further ones after using some.

To summarize,  the main loop of the algorithm is
\begin{itemize}
  \item Find the most probable correction not considered yet (one of 'triangles'),
  \item Try to expand it - decode one step or until we get to a forbidden state,
  \item Modify the tree - create new node and at most two 'triangles' - the first for the new node and the next one to the 'triangle' used.
\end{itemize}
until we get to the fixed final state on the end of the message.
\subsubsection{The weights of nodes of the correction tree}
We will work on two 'time scales' - $j$ will denote the number of states of the decoder, which corresponds to at average $J\approx H'j$ bits of the encoded message.

In each step we have standard situation for error correction: we make some observation ($O$) and we need to evaluate probabilities of its possible explanations ($E$). To cope with such problems we use Bayesian analysis, which says that probability of given situation is the probability that this explanation causes given symptoms multiplied by the probability of this explanation and normalized by the sum over all possible explanations:
\be Pr(E|O)=\frac{Pr(O|E)Pr(E)}{Pr(O)}=\frac{Pr(O|E)Pr(E)}{\sum_{E'} Pr(O|E')Pr(E')} \ee

In our case we observe some tree ($O$) and we want to find error distribution ($E$) which caused it - the most probable node to expand in this moment. We will treat $E$ as in (\ref{rtd}): it's $\{0,1\}^J$ vector in which on each position '1' denotes that we should change given bit. Finding $Pr(E)$ is easy:
\be Pr(E)=p_b^{\#\{1\leq i\leq J:E_i=1\}}(1-p_b)^{\#\{1\leq i\leq J:E_i=0\}} \label{pre}\ee
or accordingly some more complicated function if we don't assume symmetric, memoryless channel.\\

The problem is to calculate $Pr(O|E)$. We should use the real structure of the tree to find it, but it would require assuming algorithm which created it - it's becoming extremely complicated.

We will focus now on some basic method: using only the number of allowed/forbidden states outside $E$. We will see that it will already give algorithm very close to found theoretical limit for practical correction algorithms. There will be also shown some ways to improve it later.

If we assume that given correction ($E$) is proper, the states outside the corresponding path in our tree should fulfill statistics: $p_d$ of them are forbidden, $1-p_d$ are allowed:
$$ Pr(O|E)=p_d^{\#\textrm{forbidden}}(1-p_d)^{\#\textrm{allowed}} $$
Finally we can assign to each 'triangle' $Pr(O|E)Pr(E)$ - it's some multiplication of powers of $p_b,\ 1-p_b,\ p_d,\ 1-p_d$. The first pair corresponds to given correction, the second to the number of forbidden/allowed states in the rest of the tree.

Observe that the number of forbidden/allowed states outside some path is the number of all forbidden/allowed states in the tree minus these in considered path (only allowed ones). So dividing  $Pr(o|e)Pr(e)$ by the term for the whole tree, we see that we have to maximize
$$ p_b^{\#\{i:E_i=1\}}\tilde{p}_b^{\#\{i:E_i=0\}}
\tilde{p}_d^{\ -\#\ \textrm{states in this correction}}$$
among all corrections ($E$) worth to consider in this moment ('triangles'). We want to find only the node with maximal weight, so we can work on logarithm of this value. Finally while building the tree, to calculate weight for given step of decoder, we should add
\be \#\{i:E_i=1\}\cdot\lg(p_b)+\#\{i:E_i=0\}\cdot\lg(\tilde{p}_b)-\lg(\tilde{p}_d)\label{weight}\ee
to the weight of the previous step, where this time $E$ denotes bits used in the last decoding step.

We also see that using the previous algorithm - moving the 'front' of the tree, really the most probable nodes are those with the smallest number of damaged bits. The term with $\lg(\tilde{p}_d)$ allows to handle with corrections having different lengths - additionally emphasizing longer corrections.

Formally because of the search for the node with the largest weight, this algorithm has $N\lg(N)$ time complexity. In practice we usually need to work on relatively small number of nodes with the largest weight - required priority queue could have some fixed size. Very rarely it will run out and we will have to look through 'triangles' stored outside it.

\subsubsection{Analysis of correction tree algorithm with basic weights}
Let us assume that we will use this algorithm to correct some message: there is some unknown vector of errors ($E$) with $p_b$ probability of 1.

Observe that asymptotically average weight per node (\ref{weight}) is $$H'p_b \lg(p_b) + H'\tilde{p}_b\lg(\tilde{p}_b)-\lg(\tilde{p}_d)=-H'h(p_b)-\lg(\tilde{p}_d)$$
so the condition $p_d>p_d^0$ is equivalent with that the weight of the correct path is statistically growing. But error distribution is not uniform - the weight of the correct path can locally decrease. In such situation, before we will continue expanding the correct path, we have to expand some subtrees of wrong corrections.

The problem is that when local concentration of errors is very large, the weight of the correct path drops dramatically and so we have to expand very large subtrees of wrong corrections. Probability of such scenarios decreases exponentially with the size of such weight drop ($w$), but the size of such subtrees grows exponentially with it.\\

Such weight drop can have generally any length: observe that to make whole correction, from position $J$ we have to expand subtrees of wrong corrections up to weight about:
$$\min_{J'\geq 0}\left\{\#\left\{i\in [J,J+J'):e_i=1\right\}\lg(p_b)+\#\left\{i\in [J,J+J'):E_i=0\right\}\lg(\tilde{p}_b)-\frac{J'}{H'}\lg(\tilde{p}_d)\right\}$$
We need to find expected probability distribution of such drops. It doesn't depend on the position:
$$V(w):=\textrm{probability that the weight will drop by at most }w$$
For $w<0,\ V(w)=0$, but $V(0)$ corresponds to situation in which weight increases - it should be positive.

Before going to the general case, let us focus for a moment on simplified one - that all blocks have exactly 1 bit ($H'=1$). Observe that we can write equation for $V$ for two succeeding positions:
\be V(w)=\left\{\begin{array}{ll} p_b V(w+\lg(p_b)-\lg(\tilde{p}_d))+\tilde{p}_b V(w+\lg(\tilde{p}_b)-\lg(\tilde{p}_d))\quad & \textrm{for }w\geq 0\\
                 0 & \textrm{for }w<0 \end{array}\right. \label{veqv}\ee
If there would be no such boundary of behaviors in $w=0$, it would be simple functional equation - with some combination of exponents as solution. Fortunately we can use it to find the asymptotic behavior.

We know that $\lim_{w\to\infty} V(w)=1$, so let as assume that asymptotically
\be 1-V(w)\propto 2^{vw}\label{subs1}\ee
for some $v<0$. Substituting it to (\ref{veqv}), we get:
$$2^{vw}=p_b 2^{v(w+\lg(p_b)-\lg(\tilde{p}_d))}+\tilde{p}_b 2^{v(w+\lg(\tilde{p}_b)-\lg(\tilde{p}_d))}$$
\be \tilde{p}_d^v=p_b^{v+1} +\tilde{p}_b^{v+1} \label{equv}\ee
This equation has always $v=0$ solution, but for $p_d>p_d^0$ there emerges the second - negative solution we are interested in. It can be easily found numerically and simulations show that we are reaching asymptotically this behavior.\\

We can now go to the general case - we use some probability distribution of block lengths:
$$P_a:=\textrm{probability that decoding step will use }a\textrm{ bits}$$
We have $\sum_a P_a=1$ and $H'=\sum_a a P_a$.

Writing (\ref{veqv}) analogously, this time for block of length $a$ we would get $2^a$ terms. After the substitution (\ref{subs1}), we can collapse them:
$$2^{vw}=2^{vw}\sum_a P_a \left(p_b 2^{v(\lg(p_b)-\frac{1}{a}\lg(\tilde{p}_d))}+\tilde{p}_b 2^{v(\lg(\tilde{p}_b)-\frac{1}{a}\lg(\tilde{p}_d))}\right)^a$$
\be \tilde{p}_d^v=\sum_a P_a \left(p_b^{v+1} +\tilde{p}_b^{v+1}\right)^a\qquad
\left(\approx\left(p_b^{v+1} +\tilde{p}_b^{v+1}\right)^{H'}\right)\ee
Again we are interested in the $v<0$ solution. The approximation on the right is fulfilled if blocks have constant length as before, but we should also be able to use it when there are used only two block lengths differing by $1$. Generally we should be careful about it.\\

Now we have to estimate asymptotic behavior of size of subtrees of wrong correction for large weight drops. Each node of such subtree can be thought as a root of new subtree. If we will expand it for given allowed weight drop, it should give fixed expected number of nodes:
$$U(t):=\textrm{expected number of processed nodes for at most }t\textrm{ weight drop}$$
As again, $U(t)=0$ for $t<0$. For $t=0$ we process this node: $U(0)\geq 1$.

Let's focus on one bit blocks ($H'=1$) as previously. Connecting node with its children, we get:
\be U(w)=\left\{\begin{array}{ll} 1+\tilde{p}_d\left( U(w+\lg(p_b)-\lg(\tilde{p}_d))+ U(w+\lg(\tilde{p}_b)-\lg(\tilde{p}_d))\right)\quad & \textrm{for }w\geq 0\\
                 0 & \textrm{for }w<0 \end{array}\right. \label{vequ}\ee
This time we would expect that for some $u>0$ asymptotically
\be U(w)\propto 2^{uw}\ee
Substituting it to (\ref{vequ}) as previously, we will get
$$2^{uw}=\tilde{p}_d\left(2^{u(w+\lg(p_b)-\lg(\tilde{p}_d))}+
2^{u(w+\lg(\tilde{p}_b)-\lg(\tilde{p}_d))}\right)$$
\be \tilde{p}_d^{u-1}=p_b^u+\tilde{p}_b^u\ee
This equation is very similar to (\ref{equv}), for $p_d>p_d^0$ we again get two solutions. As previously, because of strong boundary conditions in $w=0$, we will be asymptotically reaching the smaller one, what confirms numerical simulations. Comparing these two equations, we surprisingly get simple correspondence between these two critical coefficients:
\be u=v+1 \ee
which is also fulfilled in the general case with different length blocks.\\

Having $U(w)$ and $V(w)$ functions, we can finally find the expected number of nodes in subtrees of wrong corrections per one node of the proper correction:
\be\int_0^\infty p_b U(w-\lg(p_b)+\lg(\tilde{p}_b))+
\tilde{p}_b U(w+\lg(p_b)-\lg(\tilde{p}_b)) \ dV(w)\ee
Because $V$ is not continuous, it's formally Stieltjes integral, but to estimate asymptotic behavior we can use $$dV(w)=\frac{dV(w)}{dw}dw\propto 2^{vw}$$
So the expected number of processed nodes per corrected bit is finite if and only if
$$ 1>2^{uw} 2^{vw}=2^{w(u+v)}= 2^{w(2v+1)}\qquad \Leftrightarrow \qquad v<-\frac{1}{2}$$
The critical $p_d$ is
\be{\tilde{p}_d}^{-1/2}=p_b^{1/2}+\tilde{p}_b^{1/2}\qquad \textrm{or generally:}\qquad
{\sqrt{\tilde{p}_d}}^{\ -1}=\sum_a P_a \left(\sqrt{p_b}+\sqrt{\tilde{p}_b}\right)^a\ee
Let us denote this value as $p_d^2$:
\be p_d^2:=1-\frac{1}{\left(\sum_a P_a (\sqrt{p_b}+\sqrt{\tilde{p}_b})^a\right)^2}\quad\qquad \left( \approx 1-\frac{1}{(\sqrt{p_b}+\sqrt{\tilde{p}_b})^{2H'}}\right)\ee
Generally $P_a$ depends on $p_d$, so we should solve this equation numerically. We will use the approximation on the right to estimate critical channel capacity for this boundary:
$$H'=H-\lg(\tilde{p}_d^2)\approx H+2H'\lg(\sqrt{p_b}+\sqrt{\tilde{p}_b})$$
$$\frac{H'}{H}\approx\frac{1}{1-2\lg(\sqrt{p_b}+\sqrt{\tilde{p}_b})}$$
It is at most 13.1\% larger (for $p_b\approx 0.03$) than for the limit for practical correction algorithms (fig. \ref{limits}). Using $p_d>p_d^2$ we can be sure that the expected width of the tree is finite - using polynomially large number of states of the decoder, asymptotically almost certainly in practically linear time this algorithm will give the proper correction.\\

This algorithm needs a bit more minimal redundancy then 'moving the front' of tree approach, because sometimes it is building huge subtrees of wrong corrections, which goes much further then the proper node. It's caused by $-\lg(\tilde{p}_d)$ term in the weight function - it amortizes large error density by the length. We see that we could improve the correction tree algorithm, by sometimes switching to the 'moving the front' algorithm, sometimes enforcing expansion of shorter paths.

For example: if the width of the tree or local error density exceeds some value, make some number of steps using only 'triangles' having position below some boundary, like this position of largest width. Unfortunately I couldn't find optimal parameters analytically, but they can be found experimentally.

\subsection{Generalized block codes}
In the previous two subsections we were using two correction mechanisms - that the final hash value (state of the coder) has to agree and that after an error in each step with probability $p_d$ we will see that something was wrong. In this subsection we will see how to use huge freedom of choice while choosing ANS coding tables to include additional error correction mechanisms known from standard block codes as in fig. \ref{conred}. This mechanism allows to immediately repair simple damages and so reduces the number of usage of decoding table, quickening the process. Finally it can be imagined as block codes in which blocks are no longer independent, but have connected redundancy. This connection is made by the internal state of the coder.

This time $p_d$ is rather large (at least 1/2), so we should use larger $H$, for example by treating a few bits as a symbol to be encoded in one step.\\

The idea is to make that Hamming distances between different allowed states are at least some fixed value. For distance 2 it can be easily done by inserting additional parity check bit, for example as the one before the oldest bit (which is always 1). In this case we can just use ScD initialization for the original symbol probability distribution and then insert parity bit, use $2l$ instead of $l$ and mark the rest (half) of states as forbidden. In this case $p_d=1/2$.

The advantage of such distance 2 code is that if among bits of one block there is only a single error, it is detected immediately. So forbidden state denotes that there was damaged one of bits used in the last step or there was at least two errors in some previous block - the set of possible correction is smaller than previously.\\

We could also enforce larger Hamming distance. Observe that e.g. triple modular redundancy codes codes can be imagined as obtained this way: $l=2^3,\ b=2$, allowed states are '1000' and '1111' - both symbols (0 and 1) has exactly one appearance. The rest of states are forbidden. While decoding step, before bit transfer, the state is always $1$, so this example is degenerated - blocks are independent.

Generally let us take $K$ as the maximum of $k_s$ for all allowed symbols (after rescaling) - so that bit transfer uses at most $K$ bits. We should enforce that for each allowed state, all states with changed at most given number of bits among these youngest $K$ positions are forbidden. In other words, if $l=2^L$, for each oldest $L+1-K$ bits we should make that two allowed states has Hamming distance at least given value - creates some block code on $K$ youngest bits.

To make the connection of redundancy work, there have to be used many essentially different block codes. We can generate them from a single one: by making XOR with some $K$ bit mask as in fig. \ref{conred} and by using some permuting on these bits - these operations make that new codewords has still the same minimal Hamming distance.

So finally to mark allowed states, for each oldest $L+1-K$ bits we should choose some $K$ bit mask to make XOR with and eventually some permutation of $K$ bits of the original block code. Then we can distribute allowed symbols among them. To make these choices, especially when we want to make encryption simultaneously, we can use PRNG as before (initialized with the key).

\end{document}